# Quantifying 'just-right' APC inactivation for colorectal cancer initiation

Meritxell Brunet Guasch[1], Nathalie Feeley[2], Ignacio Soriano[3], Steve Thorn[3], Ian Tomlinson[3,*], Michael D. Nicholson[2,*,#], Tibor Antal[1,*]

[1] School of Mathematics and Maxwell Institute for Mathematical Sciences, University of Edinburgh, Edinburgh, United Kingdom

[2] CRUK Scotland Centre, Institute of Genetics and Cancer, University of Edinburgh, Edinburgh, United Kingdom

[3] Department of Oncology, University of Oxford, Oxford, United Kingdom

* joint senior authors

[#M] for correspondence: michael.nicholson@ed.ac.uk

## Abstract

Dysregulation of the tumour suppressor gene Adenomatous Polyposis Coli (*APC*) is a canonical step in colorectal cancer development. Curiously, most colorectal tumours carry biallelic mutations that result in only partial loss of APC function, suggesting that a 'just-right' level of APC inactivation, and hence Wnt signalling, provides the optimal conditions for tumorigenesis. Mutational processes act variably across the *APC* gene, which could contribute to the bias against complete APC inactivation. Thus the selective consequences of partial APC loss are unclear. Here we propose a mathematical model to quantify the tumorigenic effect of biallelic *APC* genotypes, controlling for somatic mutational processes. Analysing sequence data from >2500 colorectal cancers, we find that *APC* genotypes resulting in partial protein function confer about 50 times higher probability of progressing to cancer compared to complete APC inactivation. The optimal inactivation level varies with anatomical location and additional mutations of Wnt pathway regulators. We use this context dependency to assess the regulatory effect of secondary Wnt drivers in combination with APC *in vivo*, and provide evidence that mutant *AMER1* combines with *APC* genotypes that lead to relatively low Wnt. The fitness landscape of APC inactivation is consistent across microsatellite unstable and *POLE*-deficient colorectal cancers and tumours in patients with Familial Adenomatous Polyposis suggesting a general 'just-right' optimum, and pointing to Wnt hyperactivation as a potential cancer vulnerability.

# Introduction

Colorectal cancer (CRC) is one of the most common and deadly cancers, with 1.9 million new cases diagnosed and 935,000 associated deaths in 2020 worldwide [1]. The Adenomatous Polyposis Coli (*APC*) gene is a canonical tumour suppressor, with loss-of-function mutations present in over 80% of sporadic CRCs [2–5]. *APC* mutations are one of the earliest, if not the earliest, genetic events in the development of CRC [6]. By dysregulating the Wnt signalling pathway, biallelic inactivation of APC in healthy colonic cells leads to the formation of adenomatous polyps, which can progress to carcinoma [7–9].

Wild type APC acts as a scaffold protein for the β-catenin destruction complex, functioning as a tumour suppressor via regulation of the Wnt pathway [10,11]. This activity involves several protein domains, including: short repeat sequences known as 20 amino-acid-repeats (20AARs), which bind to β-catenin; the β-Catenin Inhibitory Domain (CID); and the first SAMP domain, which acts as a binding site for AXIN (Figure 1a). *APC* is a classical tumour suppressor gene, requiring both alleles to be mutated for loss of function. Upon biallelic inactivation, APC loss leads to the stabilisation and accumulation of β-catenin in the cytoplasm, which, upon translocation to the nucleus, upregulates the Wnt pathway and feeds the affected cells with a permanent mitogenic signal [9,12].

Most sporadic CRCs carry mutations occurring upstream of the first SAMP repeat (codon 1569), but retain some of the 20AARs [13] [14]. A similar pattern has been observed in tumours of patients with familial adenomatous polyposis (FAP), where germline mutations removing all 20AARs are typically followed by somatic second-hit mutations that retain at least one 20AAR [15–17]. The 20AAR domains are within the large final translated exon of *APC* (codons 653-2843), in which stop-gained or frameshift mutations evade nonsense mediated decay (NMD) [18], resulting in the synthesis of truncated proteins with attenuated β-catenin binding activity [19] (Figure 1b). Progressive retention of APC regulatory repeat sequences has been associated with a successive decrease in Wnt signalling in various experimental model systems [20–23]. In particular, mutations upstream of the first 20AAR (codons 0-1256) result in maximal constitutive Wnt activity [24,25], but are rarely observed in colorectal tumours. Though the naive expectation is that complete loss of a tumour suppressor gene's function should be optimal for tumorigenesis, in most lesions APC is not fully inactivated.

These observations have led to the 'just-right' signalling hypothesis (Figure 1c), which states that both *APC* alleles are selected to retain sufficient β-catenin regulatory activity to generate an optimal Wnt signalling level for tumour growth [16,17]. In this work, we quantify this effect by analysing large cohorts of CRCs with biallelic APC inactivation, including the UK 100,000 Genomes Project [3], hereafter 100kGP, and cBioPortal cohorts (n=1,366 and n=1,305, respectively, Supplementary Tables 1 and 2). Although the genetic data is compatible with the 'just-right' hypothesis, the associations could be partly driven by mutational processes rather than selection for optimal Wnt activity. For example, genomic regions with mononucleotide repeats are particularly susceptible to insertions and deletions [26], thus the 7-base thymine repeat starting at codon 1554 in *APC* might largely explain the preponderance for partially truncated proteins. To resolve this, we propose a mathematical approach that allows us to quantify the probability of CRC progression of colonic stem cells with different *APC* genotypes, controlling for the underlying mutational processes in the colon. We quantitatively test 'just-right' against competing hypotheses, namely the 'uniform CRC risk', in which all *APC* genotypes provide the same selective advantage, and the 'maximal APC loss implies maximal risk' (Figure 1b). Furthermore, we investigate tumour heterogeneity in relation to Wnt activity based on the anatomical site of the lesion and the presence of additional mutations of secondary Wnt regulators. Finally, the generality of the 'just-right' effect is examined by comparison with hypermutant CRCs and tumours from FAP patients.

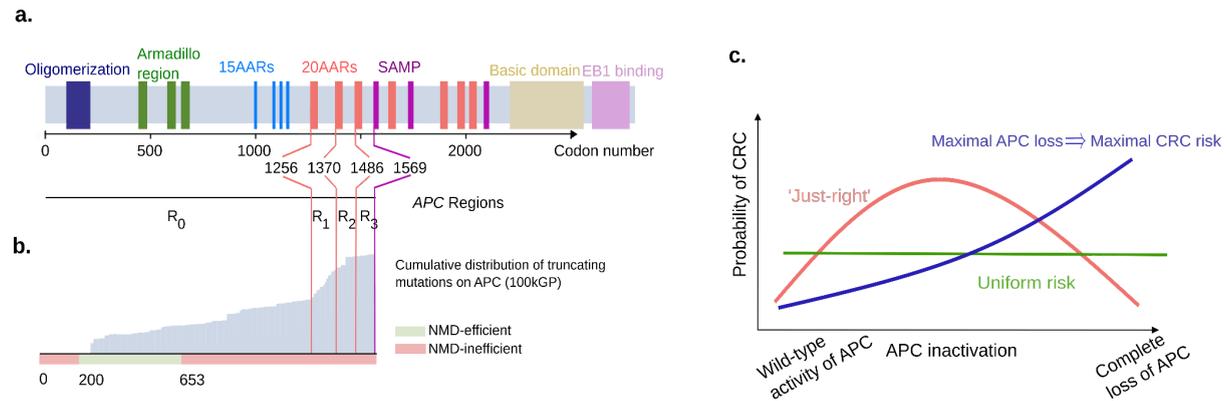

Figure 1. Evidence for 'just-right' in sporadic CRC.

(a) Schematic showing the functional domains and regions of interest of APC, and their corresponding codon position. (b) In grey, the cumulative distribution of truncating mutations of APC in the 100kGP cohort of CRCs. Below, classification of codons by efficiency of NMD. Truncating mutations affecting codons in red are expected to evade Nonsense Mediated Decay as they occur either between the start codon and upstream of the 200th nucleotide or downstream of the last exon-exon junction [18]. Notably, most truncating mutations occur downstream codon 653, which are expected to evade NMD. (c) Schematic of the 'just-right' hypothesis which posits that an intermediate level of APC inactivation maximises CRC risk, in contrast with all genotypes conferring equal risk ('uniform risk') and maximal APC loss conferring the maximal CRC risk.

## Results

**Mathematical framework to test the 'just-right' hypothesis**

We firstly profiled the distribution of mutations across the two alleles of *APC* in primary CRCs in the cohorts under study which revealed a two-dimensional mutational hotspot (Figure 2a). The hotspot suggested interdependence between the first and second hit, with most tumours retaining at least one 20AAR across both alleles. However, as discussed, the signal could be driven by mutational processes. To test and quantify the 'just-right' hypothesis for APC inactivation, and disentangle selection and mutation, we propose a mathematical framework characterising the initial stages of colorectal tumorigenesis. We first outline our mutation classification system, before detailing the mathematical model.

The *APC* genotype of colonic cells is defined by the position and class of the mutations in the two alleles. We consider all major mutation classes underlying APC inactivation:

stop-gained mutations, frameshifts, and copy number alterations separated into copy-loss of heterozygosity CL-LOH, caused by the loss of the wild-type allele, and copy-neutral loss of heterozygosity, CN-LOH, where the wild-type allele is lost and the mutated is duplicated. Since the level of APC inactivation is associated with the number of 20AAR repeat sequences retained [24,27], we classify stop-gained and frameshift mutations by regions relative to these domains and ignore somatic mutations downstream of the first SAMP repeat as they are generally not considered pathogenic, and rarely occur in tumours (Figure 1a and Figure 2a, Methods), [28]. In particular, a single truncating mutation in region $R_i$ leaves $i$ intact 20AAR repeats, where $i$ can be 0,1,2 or 3. *APC* genotypes are then denoted by (*M,N*), where *M* denotes a truncating mutation in region $R_M$ in one allele, and *N* either refers to a truncating mutation in region $R_N$ in the other allele, or it denotes a copy-loss LOH if *N*={-} or a copy-neutral LOH if *N*={x2} (Figure 2b, Methods). When there are several clonal truncating mutations, we first predict the diploid genotype of the ancestral tumour initiating cell, and then order the mutations in increasing order (*M≤N*) and only consider the two most upstream mutations. For example, genotype (1,2) refers to an *APC* genotype with two truncating mutations such that proteins synthesised from one allele will carry a single 20AAR, while proteins stemming from the other allele carry two 20AARs (see Figure 2b).

In our model, we consider how mutation accumulation in the large bowel leads to biallelic *APC* mutated cells which, in turn, can progress to cancer (Figure 2c). To disentangle mutation and selection, we first estimate the probability $m_{(M,N)}$ that a biallelic *APC* mutant cell appears with genotype (*M,N*) in the absence of selection, using mutational signature data specific to the context under consideration, e.g. signatures active in healthy colonic crypts (Methods, Table M2). We then suppose that a cell which acquired *APC* genotype (*M,N*) progresses into CRC during the patient's lifetime with probability $p_{(M,N)}$. We neglect the accumulation of further mutations in *APC* after double allelic inactivation, so-called 'third hits' [29], as these are rare in the cohorts of study (Supplementary Figure 4). Under this framework, we show (Methods) that the expected frequency of *APC* genotype (*M,N*) in colorectal cancers is given by

$$f_{(M,N)} = C\, m_{(M,N)} p_{(M,N)}$$

with *C* a positive constant independent of *APC* genotype (Figure 2c). We estimate $f_{(M,N)}$ from cohort data of sequenced primary CRCs, providing access to the *relative progression probabilities*

$$\tilde{p}_{(M,N)} = \frac{p_{(M,N)}}{\sum_{i,j} p_{(i,j)}} = \frac{f_{(M,N)}/m_{(M,N)}}{\sum_{i,j} f_{(i,j)}/m_{(i,j)}} \text{ (Equation 1),}$$

which enable assessment of the tumorigenic effect of different *APC* genotypes, while controlling for mutational processes (Figure 2d). Note that we only focus on the relative probabilities as we were unable to estimate the constant C, which conveniently cancels out in Equation 1. To relate genotypes to a measure of residual APC activity, we determine the total number *X* of 20AARs retained across the two alleles for each genotype (Figure 2b) and estimate the relative progression probability of genotypes with *X* retained 20AARs, $\tilde{p}_X$.

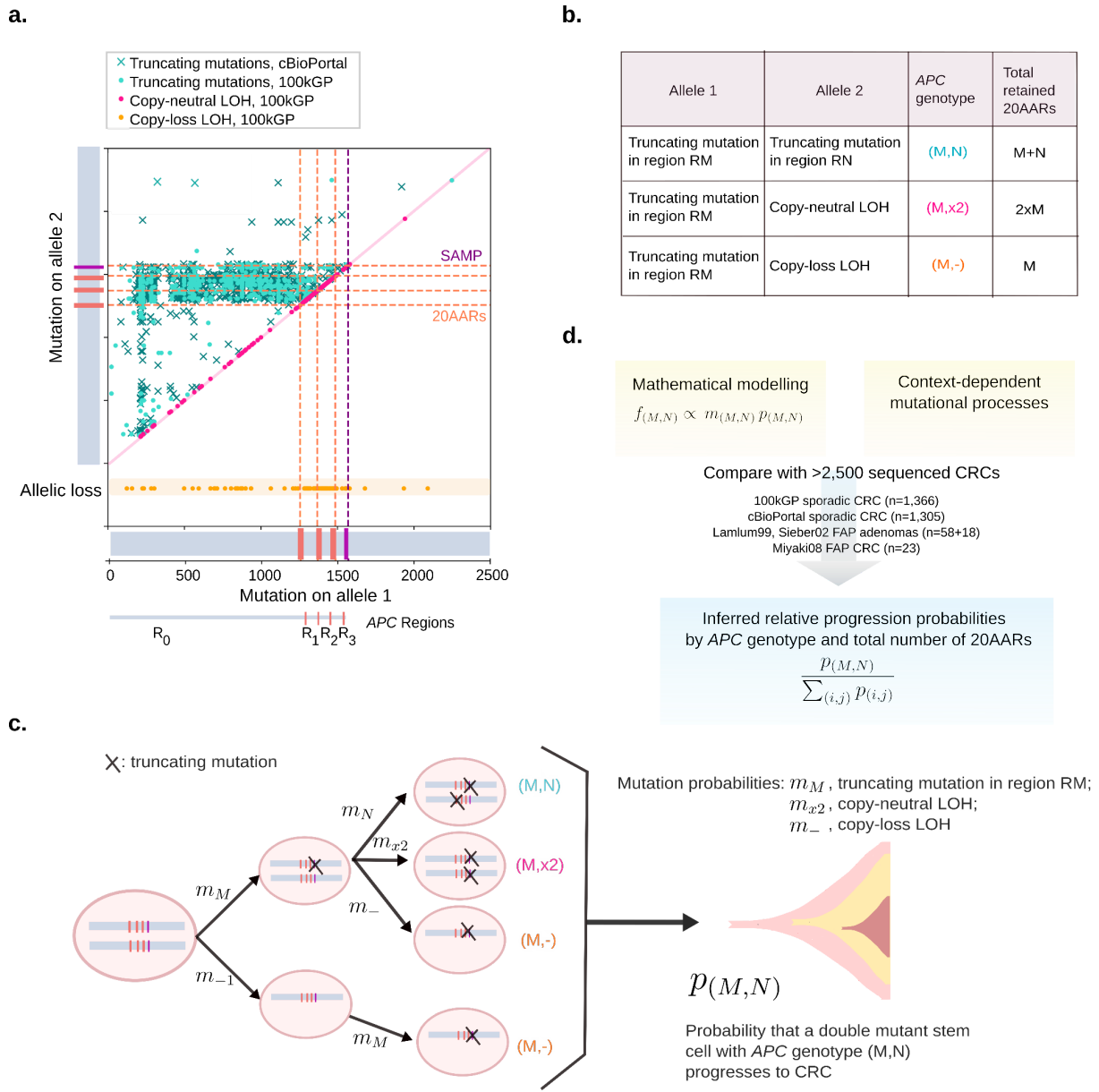

Figure 2. Mathematical approach to testing the 'just-right' hypothesis.

(a) Location of *APC* truncating mutations across cBioPortal (crosses, n=1,305) and 100kGP (dots, n=1,366) CRCs with biallelic *APC* loss. Mutation closest to 5' gene end denoted as on Allele 1 with the other mutation denoted as being on Allele 2. For cBioPortal, only tumours without copy number alterations in *APC* were considered. For 100kGP, tumours with loss of heterozygosity of *APC* via copy-neutral alteration and copy-loss of an allele, are plotted in pink and orange, respectively. The location of 20AARs and SAMP repeats is marked in dashed lines. The data displays a two-dimensional hotspot: tumours with mutations in region $R_0$ of allele 1 tend to have mutations between regions $R_1$ and $R_2$ of allele 2, and points to the 20AARs limiting the regions of interest. (b) Classification of biallelic *APC* mutant cells by the position and class of the two hits, and the corresponding total number $X$ of 20AARs retained across the two alleles. (c) Mathematical model of CRC initiation, in which cells accumulate truncating mutations of *APC* in region $R_M$ with probability $m_M$, copy-loss LOH with probability $m_-$ or copy-neutral LOH with probability $m_{x2}$. Once a stem cell has lost both copies of *APC*, the cell progresses into cancer with a probability that depends on the *APC* genotype, $p_{(M,N)}$. From the model, the expected frequency of cancers with a given genotype, $f_{(M,N)}$, can be derived, which is comparable to cancer sequencing data. (d) Schematic of the strategy developed to infer the relative probability of progression of genotype (*M,N*), $\tilde{p}_{(M,N)}$, by combining mathematical modelling with sequence data from sporadic and familial *APC*-driven CRC.

## 'Just-right' APC inactivation for CRC initiation

To test and quantify the effect of different APC inactivation levels for cancer initiation, we applied our mathematical model to the 100kGP cohort [3]. Initially, we considered microsatellite-stable (MSS) primary tumours with double allelic inactivation of APC, and without pathogenic mutations of DNA polymerase epsilon (*POLE*) (n=1,037, filtering details are in Methods).

First, we parametrized the model using mutational signatures active in healthy colonic crypts [30] to estimate the probabilities of truncating mutations in different regions of *APC* under neutral evolution (Figure 3a-c). Primarily due to $R_0$ being the longest region, the majority of variants are expected to occur within $R_0$ (Figure 3a). However frameshift mutations are relatively biassed to $R_3$, due to the activity of indel signature ID2 acting on a 7-base thymine mononucleotide repeat starting at codon 1554 (Figure 3b). Assuming that genotypes retaining 0 copies of 20AARs are equally tumorigenic, the proportion of those that have copy-number alterations is informative of the relative rates of CL-LOH and CN-LOH compared to single base substitutions (SBS). By using a SBS rate in healthy colonic crypts of $1.45*10^{-8}$ [30], and noting that CN-LOH can only occur as a second hit, we find rates of $4.72*10^{-6}$/cell/year for *APC* CL-LOH and $7.18*10^{-6}$/cell/year for *APC* CN-LOH (Methods).

Combining the mutation probability estimates of different *APC* genotypes with the corresponding frequencies in MSS CRCs in the 100kGP cohort, we estimated the relative progression probabilities of *APC* genotypes, $\tilde{p}_{(M,N)}$. Remarkably, we found that genotypes (1,1), (1,x2), (0, 2) and (2,-) have around 50 times higher progression probabilities than genotype (0,0) (Figure 3d, Supplementary Table 7). Thus, we rejected the 'uniform CRC risk' hypothesis that all genotypes have the same cancer progression risk, showing that selective pressures shape the *APC* genotype distribution (non-overlapping 95% CIs).

The total number of 20AARs explains a considerable degree of variability in the relative progression probabilities between genotypes ($R^2$=0.82), supporting a model in which the total number of 20AARs across both alleles determines APC activity. This can be directly observed in Figure 3d, which shows a striking concordance in the relative progression probabilities amongst genotypes that result in the same total 20AARs. We reject the hypothesis that maximal loss of APC provides maximal CRC risk (0 not in 95% CI of mode), instead finding that an intermediate level of APC loss with a total of 2 copies of 20AARs results in maximal tumorigenic effect. Remarkably, the relative progression probability of 0 copies of 20AARs, which corresponds to maximal APC loss, is similar to that of retaining 6 copies of 20AARs, which is thought to retain most APC activity.

Similar results were found when analysing primary MSS CRCs in the cBioPortal cohort [4,5], where we considered biallelic diploid *APC* mutant samples [4,5]. We again reject both the 'uniform CRC risk' and the 'max-loss implies max-risk' hypotheses (95% CI, bootstrapping), and find that the total number of 20AARs explains variability between genotypes ($R^2$=0.89, Supplementary Figure 1). In Figure 3e, we plot the relative progression probabilities by the number of 20AARs in the 100kGP and cBioPortal cohort, showing almost identical curves across the two independent cohorts when classifying tumours by the total number of retained 20AARs. Since our analysis indicates that the total number of 20AARs explains most genotypic variability, in the rest of this work, we focus on the progression probabilities of cells with different total numbers of 20AARs retained.

Given the correlation between APC inactivation and Wnt activity [20–23,31], the above findings support the hypothesis that a just-right level of Wnt dysregulation leads to maximal cancer risk. However, a considerable proportion of tumours develop through 'non-optimal' APC inactivation levels (e.g. 14.5% of tumours in 100kGP retain 0 copies of 20AARs). Next, we study other factors that influence Wnt activity to understand the variability in CRC progression risk amongst lesions with the same *APC* genotypes.

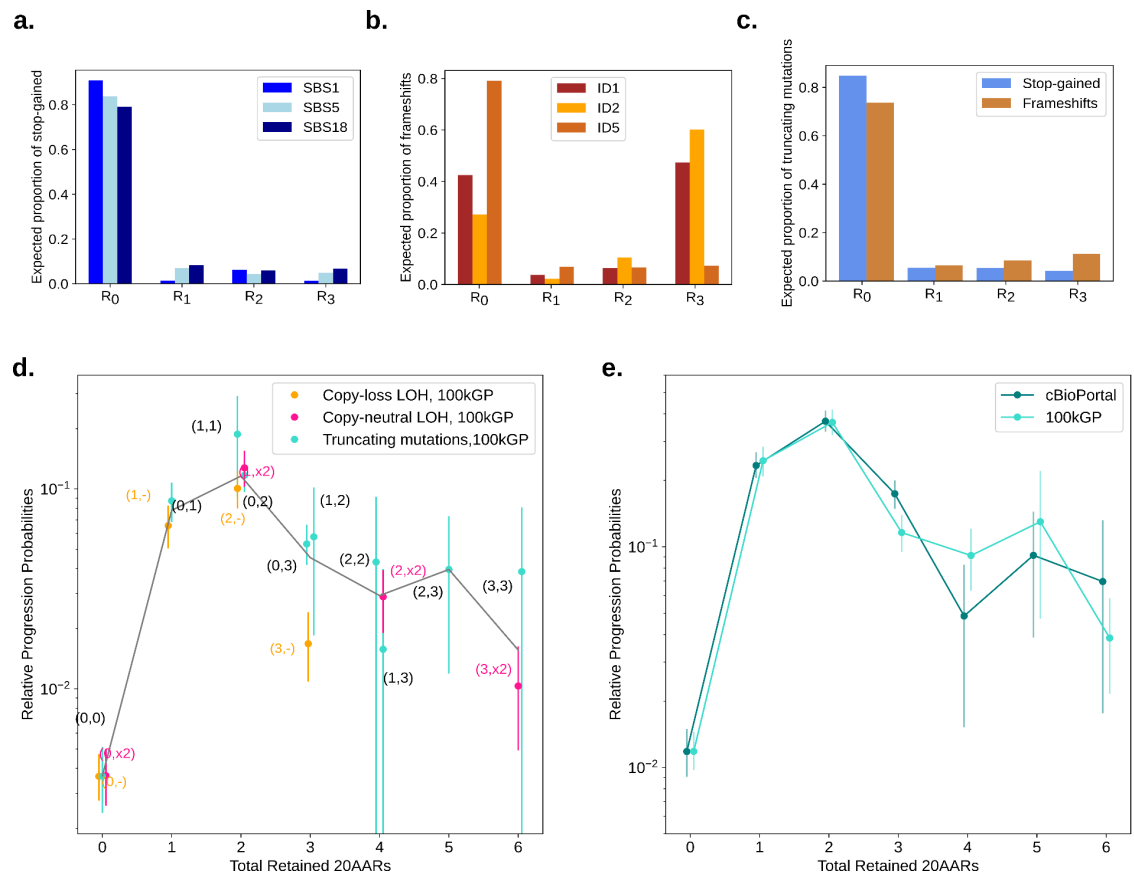

Figure 3. Optimal number of 20AARs for CRC progression .

(a, b) The proportion of stop-gained and indels, respectively, expected to fall in different regions of *APC*, estimated by considering the ubiquitous mutational signatures found in healthy colon crypts [30]. (c) The expected proportion of truncating mutations in each region which is used to estimate the rates of truncating mutations in each region. (d) The relative progression probability of different *APC* biallelic genotypes, $\widetilde{p}_{(M,N)}$, is plotted against the total number of 20AARs retained across both alleles.

The frequencies of genotypes were calculated from sequence data of MSS primary CRCs in the 100kGP cohort (n=1,037, Methods). Whiskers represent 95% confidence intervals (bootstrapping). The grey line is the average of the progression probability over all genotypes resulting in a given number of retained 20AARs, weighted by the number of samples. (e) The relative progression probability of different total number $X$ of 20AARs retained across both alleles of *APC*, $\tilde{p}_X$, with frequencies calculated from sequence data of MSS primary CRCs in 100kGP (n=1,037, Methods) and cBioPortal (n=1,041, Methods).

## APC inactivation varies across anatomical sites

Molecular differences between lesions in different colonic sites have been identified [27,32], which could contribute to variability in 'just-right' Wnt levels. Mutational signature burden differs significantly across anatomical locations in both healthy crypts [30] and CRCs [3,27]. However, we found that the difference in signature proportion was relatively minor in healthy crypts (Supplementary Figure 2, Supplementary Table 5), hinting that site-specific mutational processes are unlikely to play a major role in location-specific *APC* genotype patterns. To isolate-out variability in selection, we accounted for site-specific mutational processes, and calculated the relative progression probability curves separately for both proximal and distal (including rectum) CRCs, finding significant differences between anatomical sites (Figure 4a).

To quantify differences in selection and relate them to Wnt activity, we computed the progression-weighted mean 20AARs number, defined as the average number of 20AARs weighted by the corresponding progression probabilities (Methods, Equation M5). This can be interpreted as a proxy for the optimal level of Wnt activation contributed by APC loss. We can then compare two subtypes of cancers A and B by calculating the difference $\Delta_{A\text{-}B}$ between their respective progression-weighted mean 20AAR number (Methods, Equation M6). If $\Delta_{A\text{-}B} > 0$, the shift suggests that tumours in subtype A "prefer" to retain a higher number of 20AARs and thus require lower Wnt activity, and vice versa for $\Delta_{A\text{-}B} < 0$.

Using the $\Delta$ measure, we found a significant difference when stratifying tumours by anatomical site, with the progression-weighted mean 20AARs number being higher amongst proximal tumours compared to distal ($\Delta_{\text{Proximal-Distal}}=1.1$, p<0.001, permutation test) (Figure 4b). This suggests that tumours in the proximal colon benefit from lower Wnt activation due to APC loss. We considered other clinical features reported in the 100kGP cohort that could underlie variability of *APC* genotypes, but found no significant differences (permutation test, p>0.05) upon stratifying by sex or early onset cancers, defined as <50 years old at resection (Figure 4B, Supplementary Figure 3).

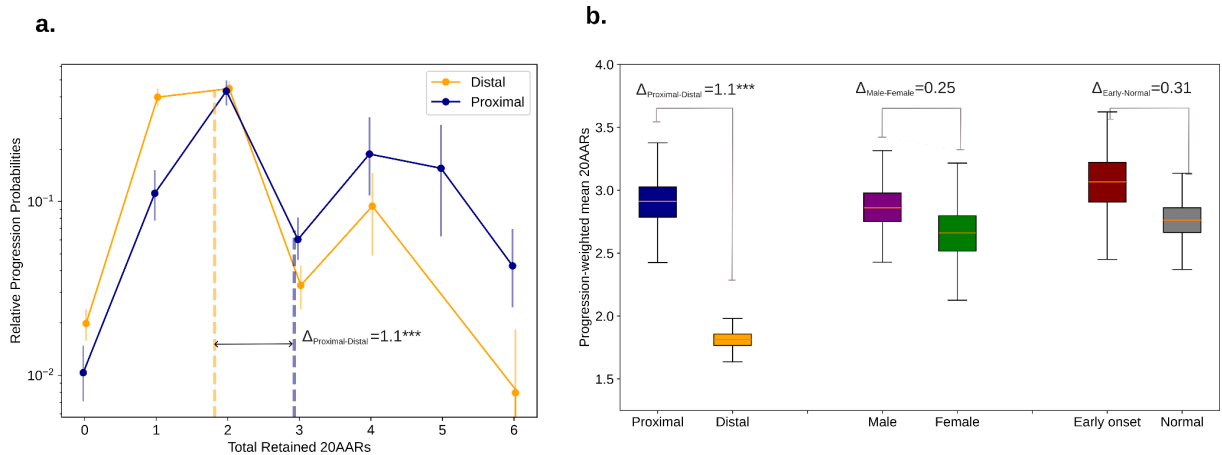

Figure 4. 'Just-right' APC inactivation is higher in the distal colon.

(a) The relative progression probability versus total number of 20AARs retained over both alleles, controlling for site-specific mutational processes, for proximal (blue) and distal (orange) cancers, with genotype frequencies calculated from bulk sequence data of MSS primary CRCs in the 100kGP cohort (n=313 proximal, n=574 distal/rectum). Whiskers on points indicate 95% confidence intervals (bootstrapping). Thick dashed vertical lines indicate the progression-weighted mean 20AARs number retained, representing the optimal level of Wnt activation contributed by APC loss. Proximal tumours are under selection for a higher number of 20AARs. (b) The progression-weighted mean 20AARs number retained in different tumour stratifications, whiskers on points indicate 95% confidence intervals (bootstrapping). We find a significant difference of $\Delta_{P-D}$=1.1 between proximal and distal tumours (p<0.001, permutation test), but no statistically significant differences between tumours in male versus female patients (p=0.25, permutation test), nor in patients with early onset (<50 years old at resection) versus normal onset (>50 years old at resection) (p=0.31, permutation test).

## Secondary Wnt drivers can combine with APC inactivation to achieve 'just-right' Wnt signalling

While *APC* is the main Wnt driver in CRC, other genes are also thought to dysregulate the Wnt pathway when mutated [33,34]. Thus, we next investigated tumours with additional mutations in Wnt drivers to study the 'just-right' hypothesis at the pathway level. Primary Wnt drivers, such as inactivation of *RNF43,* activation of *CTNNB1,* or *RSPO* fusions, have similarly drastic effects as APC inactivation on Wnt [35–37]. In the 100kGP cohort, alterations of *RNF43* or *CTNNB1* are found in a minority of sporadic CRCs, mostly microsatellite unstable (MSI), and are mutually exclusive with APC inactivation in MSS tumours (OR=0.019, p=3.94 $10^{-24}$ and OR=0.15, p=1.46 $10^{-5}$ respectively, Figure 5a, Supplementary Table 8), suggesting an upper bound on Wnt activity. Other Wnt drivers with smaller effects on Wnt activity can co-occur with primary Wnt drivers - these are referred to as secondary Wnt drivers [38,39]. In 100kGP, driver mutations of *AMER1*, *SOX9* and *TCF7L2* co-occur with *APC* in MSS tumours (OR=15.53, p=2.34 $10^{-5}$; OR=2.62, p=7.20 $10^{-4}$ and OR=2.35, p=1.47 $10^{-3}$, respectively, Figure 5A, Supplementary Table 8). However, their directional effect, that is whether they increase or decrease Wnt activity, and their role in 'just-right' signalling, remain unclear.

By comparing MSS CRCs with and without secondary Wnt driver mutations, we reasoned that the effect of the secondary Wnt drivers could be measured under the following rationale. Assuming the 'just-right' model for Wnt activity, secondary Wnt driver mutations that cause increased Wnt are expected to be more frequent in combination with *APC* genotypes that lead to relatively low Wnt, i.e. those that retain more 20AARs, resulting in a rightward shift in the relative progression probability curve, $\Delta_{mutant-WT}>0$ (Figure 5b). Similarly, secondary drivers that cause reduced Wnt would be more common with Wnt high *APC* genotypes, which retain fewer 20AARs, hence a left-ward shift in the relative risk curve is expected, with $\Delta_{mutant-WT}<0$ (Figure 5b).

In agreement with the theoretical expectation, a clear shift to the right was observed in tumours with driver (loss of function) mutations in *AMER1* (Figure 5c). This shift indicates that *AMER1* mutations tend to occur in tumours with lower than average APC inactivation, potentially increasing Wnt activity to the 'just-right' window, in accordance with both *in vitro* and *in vivo* experiments showing that wild type AMER1 reduces Wnt signalling [40–42]. Conversely, a shift to the left was observed in tumours with driver mutations in *TCF7L2* (Figure 5c), which all retain 0-3 copies of 20AARs. To quantitatively classify genes into Wnt up or down-regulators, we computed $\Delta_{mutant-WT}$ weighted by the proportion of mutations occurring in proximal or distal tumours, thus obtaining a metric that is independent of site-specific biases (Methods). Using this measure, we predicted mutated *AMER1* and *SOX9* as Wnt up-regulators ($\Delta_{mutant-WT}>0$, 95% CI, bootstrapping), and mutations of *TCF7L2* and *BCL9L,* as Wnt downregulators ($\Delta_{mutant-WT}<0$, 95% CI, bootstrapping) in tumours with *APC* inactivation, relative to the wild type protein. Considering that most driver mutations in the genes above result in loss of function [3] (Supplementary Table 9), the findings are consistent with current understanding of the wild-type proteins functions, e.g. AMER1 and SOX9 promote APC activity, acting as Wnt repressors in healthy tissue, whilst TCF7L2 and BCL9L promote β-catenin transcription [43]. Mutations of FBXW7 and BCL9 were consistent with no effect on the cancer progression risk of APC genotypes, although this may be due to low sample size, and variant-specific functional consequences. For *AXIN1*, *AXIN2* and JUN, site-correction was not possible due to the limited number of samples.

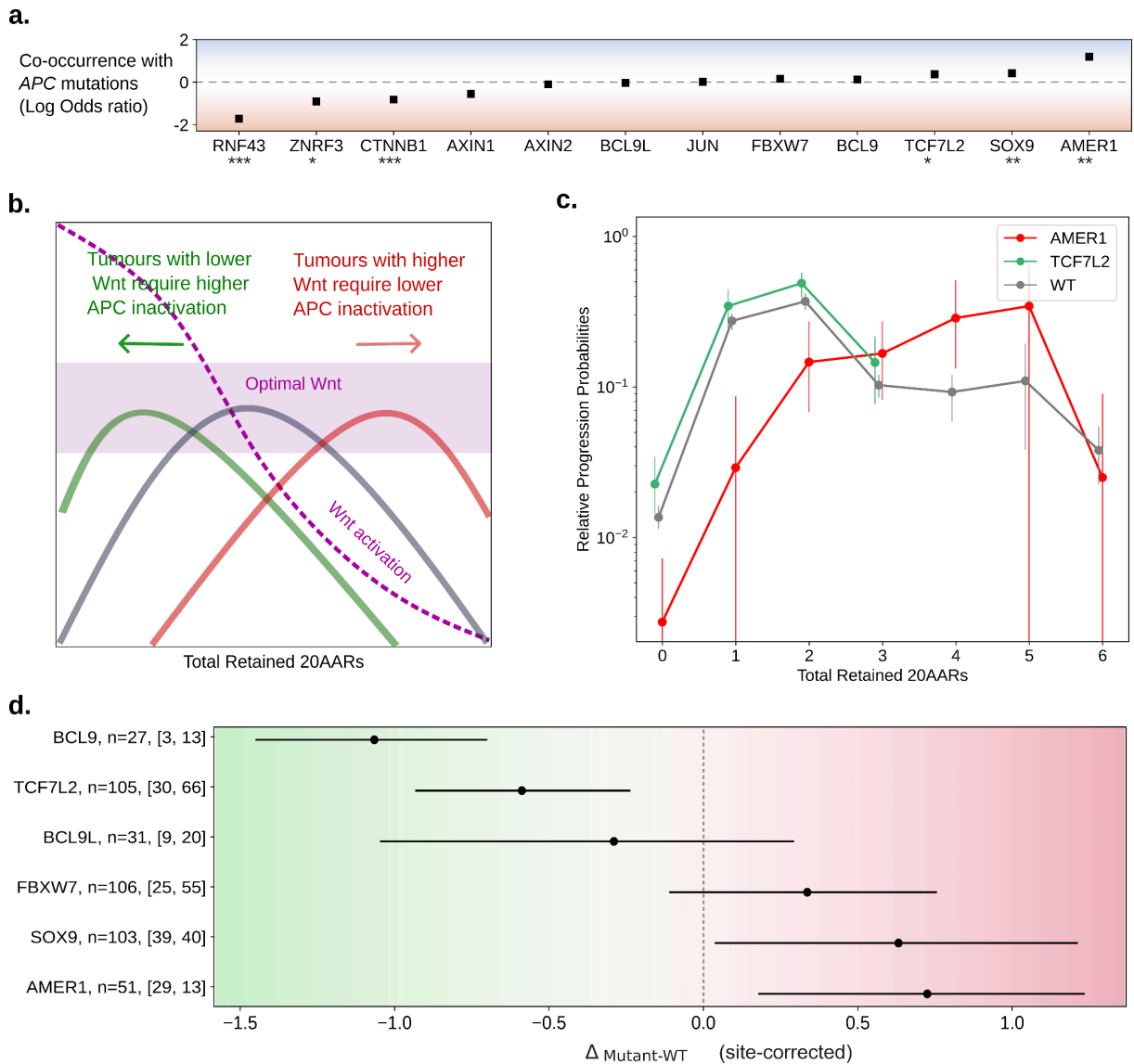

Figure 5. 'Just-right' Wnt activity at the pathway level.

(a) Odds-ratio between APC inactivation and pathogenic mutations in other Wnt related genes, in the 100kGP MSS cohort (n=1,639, Supplementary Table 8, Fisher's test,* p<0.05, ** p<0.01,*** p<0.001). (b) Schematic of the effect of additional mutations in Wnt pathway regulators. Assuming that the cancer progression probabilities of *APC* mutant cells are due to the corresponding level of Wnt pathway activation, tumours with Wnt upregulating mutations will require a smaller Wnt contribution from *APC* mutations, and so will have relative progression probability curves shifted to the right, and vice-versa. (c) Relative progression probabilities as a function of the total number of retained 20AARs, using sequence data of MSS primary CRCs with pathogenic *AMER1* mutations (n=51, red), with *TCF7L2* mutations (n=105, green), and tumors without mutations in non-*APC* Wnt regulators (n=825 grey). Whiskers for 95% CI (bootstrapping), thick dashed vertical lines indicate the progression-weighted mean 20AARs number retained. (d) Difference in progression-weighted mean 20AARs number for tumours with pathogenic mutations in different Wnt genes, $\Delta_{mutant-WT}$, corrected by the effect of anatomical site (Methods). Horizontal bars for 95% confidence intervals (bootstrapping). Numbers next to the gene labels indicate the total number of tumours with mutations in the Wnt driver, and the number of which were classified as proximal and distal colon, respectively.

## APC inactivation in hypermutant tumours

Thus far we have focused our analysis on MSS CRCs, and excluded hypermutant CRCs - that is CRCs with mutations affecting the proofreading capability of DNA polymerase epsilon (*POLE*), and microsatellite unstable (MSI) CRCs. These tumours not only have an increased mutational burden, but are also characterised by distinct mutational processes [44,45]. Thus, it is not surprising that the landscape of *APC* mutations in *POLE* and MSI CRCs in the 100kGP cohort differs from MSS CRCs (Figure 6a-c). To assess whether hypermutant CRCs comply with the 'just-right' distribution observed for MSS cancers (Figure 4), we first studied how the intrinsic mutational processes active in hypermutant cancers affect the distribution of *APC* genotypes, assuming that *POLE* mutations and mismatch repair deficiency precedes APC inactivation [46] [47].

Integrating data on *POLE* mutational signatures [48] with the sequence context of *APC* (Methods), we found that *POLE*-mutant associated signatures result in an expected increased proportion of stop-gained mutations in *APC* regions $R_1$ and $R_3$ compared to healthy crypts (Figure 6d, Supplementary Table 5). Since no frameshifts in *APC* were observed in the *POLE* CRCs in 100kGP (Figure 6b), we omitted the indel analysis for these tumours. To analyse the distribution of *APC* genotypes in lesions with microsatellite instability (MSI), we used the genome-wide mutational signatures found in >85% of MSI CRCs in the 100kGP cohort [3], n=364). Notably, the combined MSI indel signature results in a 5-fold bias for frameshifts in region $R_3$ compared to MSS (Figure 6d, Supplementary Table 5). This bias can be observed in Figure 6c, which displays a sharp increase in the number of frameshift mutations in region $R_3$, and should lead to more retained 20AARs in MSI compared to MSS. As we have shown that proximal lesions are more likely to progress if they retain more 20AARs, the bias could explain, in part, why *APC*-driven MSI lesions tend to occur relatively often in the proximal colon compared to MSS (63:22 and 313:574 proximal:distal ratios in the 100kGP cohort, respectively).

We analysed the distribution of *APC* mutations in CRCs in 100kGP with pathogenic *POLE* mutations or MSI, mirroring the analysis carried out for MSS CRCs (Methods). We excluded lesions with copy-number alterations, resulting in n=17 *POLE* samples and n=64 MSI CRCs. While subtle differences in the relative progression probability curves were observed (Figure 6e, Supplementary Figure 5), we again reject the hypothesis that complete APC loss provides maximal CRC risk, with 2 copies of 20AARs providing maximal risk in both in POLE-deficient and MSI CRCs (2 20AARs in the 95% CI, bootstrapping). Notably, we found no significant differences in the progression-weighted mean 20AARs number compared to MSS tumours ($\Delta_{MSS-POLE}$=-0.22, 95% CI=[-0.62, 0.15], $\Delta_{MSS-MSI}$=-0.29, 95% CI=[-0.67, 0.08], bootstrapping), whilst, without adequately correcting for the characteristic mutational signatures of *POLE* and MSI, the differences were larger, and statistically significant in the case of MSI tumours ($\Delta_{MSS-POLE,nc}$=-0.31, 95% CI=[-0.78, 0.2], $\Delta_{MSS-MSI,nc}$=-1.87, 95% CI=[-2.51, -0.81], Supplementary Figure 6). This finding emphasises the importance of mutational bias analysis, and suggests that repair deficiencies are indeed required prior to APC inactivation.

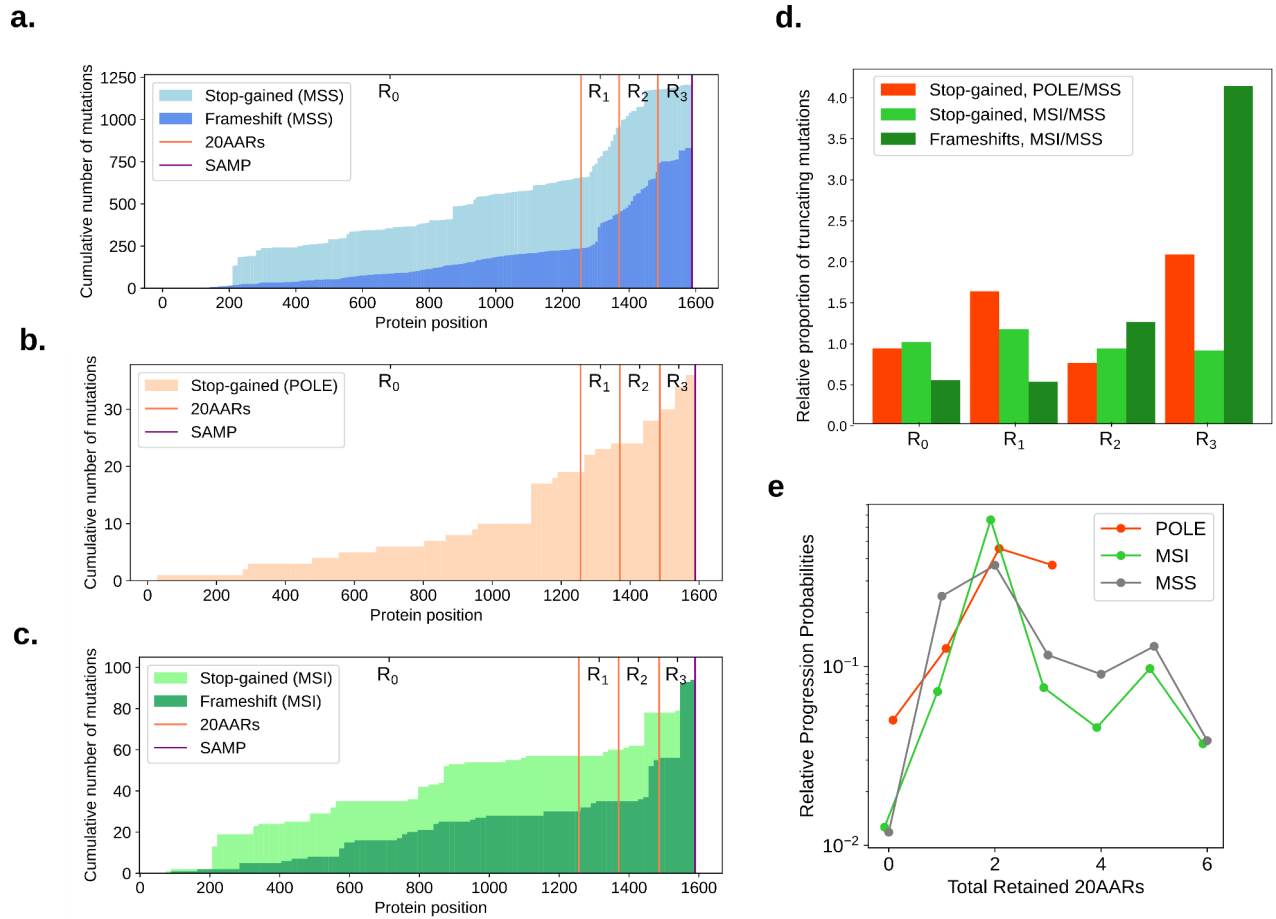

Figure 6. 'Just-right' APC inactivation in POLE and MSI CRCs.

(a-c) Cumulative number of stop-gained and frameshift mutations detected per codon position of *APC* in MSS (a), *POLE*-mutant (b) and MSI (c) primary CRCs in the 100kGP cohort. Vertical lines indicate the locations of the 20AAR domains and the SAMP repeat. (d) Expected proportion of mutations of different regions of *APC* in *POLE*-mutant and MSI relative to MSS, calculated using the mutational signatures detected in healthy colonic crypts [30], *POLE*-mutant crypts [48], and MSI colorectal cancers [3] (Supplementary Tables 3-5). (e) The relative progression probabilities by total number of 20AARs retained in MSS, *POLE*-mutant and MSI CRCs in the 100kGP cohort.

## 'Just-right' APC mutations in FAP

Evidence for the 'just-right' hypothesis for *APC* mutations initially came from familial adenomatous polyposis (FAP) patients, carrying germline mutations of *APC*. FAP patients present with large numbers of polyps at early ages, and in most cases develop CRC unless treated. In agreement with the 'just-right' hypothesis, the somatic hit depends on the germline *APC* mutation, with most lesions retaining 1-3 copies of 20AARs [16,17,49]. Moreover, FAP patients show variable polyp burdens and time of cancer onset depending on the specific germline mutation [16,17,49].

We used the progression probabilities inferred from the analysis of sporadic MSS CRC to assess the concordance between sporadic and FAP tumours. In Figure 7, the expected distribution of the somatic *APC* hit is compared to the observed distribution in FAP patients from public datasets [15], 86 CRCs from 23 FAP patients; [16], 92 adenomas from 5 FAP patients). For FAP patients with germline mutations in regions $R_1$, $R_2$ and $R_3$, the FAP data is consistent with the expected distribution, with most adenomas and CRCs developing via LOH or mutations of region $R_0$, (Figure 7b-d). However, for patients with germline mutations in $R_0$, the sporadic CRC distribution underestimates the proportion of lesions that develop via mutations in $R_2$, whilst the proportion of tumours with LOH is underestimated in patients with germline in $R_1$ or $R_2$. The discordance might exist for several reasons, including: polyclonality of some polyps in FAP; unmeasured factors in the FAP data including anatomical site of the lesion or mutations in secondary Wnt drivers, which both affect optimal *APC* genotypes as discussed above; different selective pressures might exist in FAP patients further enhancing the selective pressure for 1-2 retained 20AARs, e.g. due to intercrypt competition. Broadly, the concordance between the predicted and observed distribution of the somatic hit suggests similar selection for intermediate APC inactivation in FAP tumours.

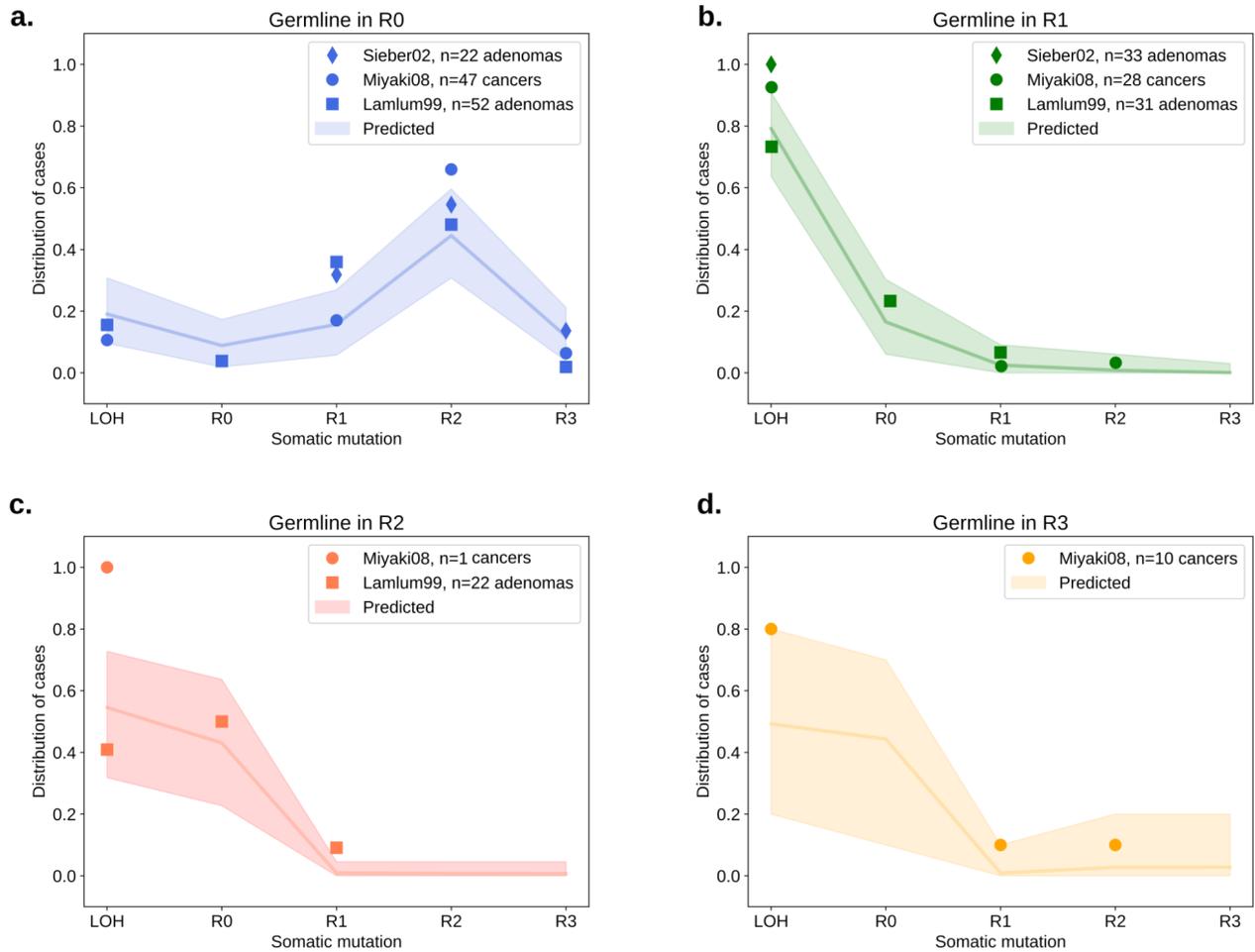

Figure 7. 'Just-right' in FAP patients.

(a-d) The distribution of the *APC* somatic hit on tumours in FAP patients with germline mutations in different regions. Points indicate the observed distribution in FAP patients from different studies [15,16]. The line indicates the expected distribution calculated using the mutation and cancer progression probabilities estimated from healthy crypts and sporadic CRC data, with shaded regions for a conservative 95% CI, obtained by performing multinomial simulations with number of trials given by the maximal number of patients across the studies in each germline group.

## Discussion

Repression of APC function is the canonical tumorigenic event in colorectal cancer, resulting in the Wnt pathway dysregulation that is a pervasive feature of this cancer type. The tumour suppressor activity of APC relies crucially on its 20 amino acid repeat domains (20AARs), which bind to β-catenin. By integrating somatic and cancer datasets with mathematical modelling, correcting for mutational bias, we find a quantitatively consistent signal that biallelic *APC* genotypes that retain intermediate β-catenin binding activity confer maximal tumorigenic effect in MSS and hypermutant sporadic CRCs, as well as tumours in FAP patients. While a degree of variability in the fitness conferred by specific mutations of oncogenes is expected, e.g. G12D variants compared to A146T in *KRAS* [50], that complete loss of the β-catenin-binding 20AARs in tumour suppressive APC does not lead to maximal cancer risk is remarkable.

Although previous experimental work suggests differential β-catenin binding strengths amongst the different 20AAR domains [24], our findings indicate that genotypes with the same total number of 20AARs retained have similar CRC progression risks, independent of the mutational pathways. For MSS CRCs, we found that cells with biallelic *APC* genotypes resulting in 2 retained 20AARs are at 50 times higher probability of progressing to CRC than those in which all binding domains are lost. As the number of repeats retained is inversely correlated with Wnt activation (Kohler et al. 2008; Kohler et al. 2009), this suggests that *APC* mutations are selected to result in intermediate Wnt activity, pointing to hyperactivation of Wnt signalling as toxic for tumour development, in concordance with recent experimental studies [51]. However, we note that variability amongst genotypes could also be related to non-Wnt related essential functions of APC (Zhang and Shay 2017; Hankey, Franke, and Groden 2018).

By assuming equivalent tumorigenic effect between point mutation and copy number driven biallelic loss of all 20AARs, we estimated the rates of CL-LOH and CN-LOH driving APC inactivation in the healthy colon as $4.72*10^{-6}$ and $7.18*10^{-6}$/cell/year, respectively. These are around an order of magnitude lower than previous estimates ([52], Supplementary Information), yet in better agreement with the ratio of LOH at the *APC* locus in CRCs in the 100kGP cohort and the FAP patients analysed (Figure 7).

Quantitatively strengthening prior observations [27], we found that proximal and distal tumours are under selection for different levels of APC inactivation, with an effect independent of site-specific mutational processes. Proximal tumours retain a higher number of 20AARs, and thus appear to require lower activation of the Wnt pathway, agreeing with prior murine studies [32]. This might be partly explained by a higher baseline expression of Wnt genes in the proximal colon, as reported in expression assays in murine and human colonocytes [32,53]. In addition, it might indicate that the 'just-right' window varies across the colon, with lesions in the distal under selection for higher Wnt, e.g. due to enhanced immune surveillance [54].

We found significant differences in the *APC* relative progression curves in tumours with mutations in secondary Wnt drivers. In particular, we found that pathogenic mutations of *AMER1*, which are thought to upregulate the Wnt pathway [40–42], are associated with *APC* mutations that result in lower Wnt activation. In line with the mini-driver model [39,55], our findings suggest that secondary Wnt drivers combine with different biallelic *APC* genotypes to achieve 'just-right' Wnt. Moreover, our analysis provides a framework to estimate the effect on Wnt activity of secondary Wnt drivers that co-occur with APC inactivation, indicating that, on aggregate, mutations in *SOX9* and *AMER1* are Wnt-upregulators, while mutations in *TCF7L2* and *BCL9L* down regulate Wnt.

Despite its name, motivated by the initial discoveries [16,56], the 'just-right' model should not be interpreted entirely deterministically, in the sense that only specific *APC* mutations can drive tumorigenesis. Instead, we suggest that 'not-right' mutations might still lead to cancer progression, albeit with a much lower probability. Isolating precisely when the 'just-right' selective pressure acts during oncogenesis, e.g. enabling dysplastic polyp formation versus progression from adenoma to carcinoma, will require comparative analyses with mutation data from pre-cancerous tissues. Beyond fundamental understanding, recent efforts to exploit hyperactivation of cancer pathways for therapeutics [51] motivate the need to quantify the mutation-specific levels of pathway dysregulation. Exemplified by our analysis of *APC* genotypes within the context of secondary Wnt driver mutations, cancer genotypes should be assessed within the context of their associated signalling pathways to incorporate epistatic effects, as opposed to adopting solely a mutation-centric lens. With a growing amount of available data on somatic mutational processes and tumour mutational landscapes, pathway informed computational approaches as proposed in this work are necessary to quantify how the forces of selection and mutation combine to shape cancer evolution.

# Methods

## M1. Classification of APC-driven CRC tumours

*1.1 Genomics England*

We used cohort data of version 5 of the UK 100,000 Genomes Project, which performed whole genome sequencing on 2,023 paired cancer (~100x average depth) and normal (blood, 33x) samples from 2,017 CRC patients (median age at sampling 69 years, range 23-94; 59.4% male). This is the same cohort used by Cornish *et al.* [3].

We considered primary cancers with somatic pathogenic mutations in *APC* (n=1,376, 84.22%), and excluded patients that had received radiotherapy prior or other treatment to surgery, or had any germline pathogenic mutations in *APC* or Lynch Syndrome genes. Unless otherwise stated, the analysis was performed on cancers with MSS status and no pathogenic *POLE* mutations (n=1,263). Splicing mutations were excluded, as well as any missense mutations, and stop-gained and frameshift mutations occurring downstream of the SAMP repeat (amino acid position 1569). We also excluded samples with a single *APC* mutation and no evidence for copy number alterations (n=140).

In order to determine the *APC* genotype in terms of 20AARs of each sample, we first identified the position of stop-gained and frameshift mutations. In particular, we classified stop-gained and frameshift mutations in four regions of interest, relative to the 20AARs domained:

| Region | Amino acid position |
|---|---|
| $R_0$ | [0, 1256] |
| $R_1$ | (256, 1370] |
| $R_2$ | (1370, 1486] |
| $R_3$ | (1486, 1569] |
| Table M1. Classification of *APC* into regions. | |

**Mapping to *APC* genotype at initiation.** In order to classify samples by the copy-number of *APC* at tumour initiation, we used the allele-specific copy number alteration calls of Cornish *et al.* [3] at the *APC* site of chromosome 5 (5q22.1–q22.3), as well as whole genome duplication (WGD) status estimated by Conrish *et al.* [3]. We denote the allele-specific copy-number of a sample by [a, b], where 'a' gives the relative contribution to sequence reads coming from the major allele, and 'b' from the minor allele. In keeping with prior observations [57] whole-genome duplications were assumed to occur after APC inactivation. Thus, for tumours with negative WGD status, we classified them as normal diploid if the copy number was [1, 1]; as CL-LOH if they were [1, 0] and CN-LOH if [a, 0] for a≥2 (assuming a CN-LOH event was followed by amplification for a>2). For tumours with positive WGD status, we classified [a, b] as normal diploid at initiation with a,b>0, whilst [a, 0] was classified as CL-LOH for a=1,2, and CN-LOH for a>2. Samples with other copy-number alterations were excluded, as well as samples with a single truncating mutation upstream SAMP and no copy-number alterations. Supplementary Table 6 summarises the mapping to *APC* genotype at initiation for all considered combinations of copy number, WGD and annotated variants.

**Classification by number of retained 20AARs.** Given the *APC* genotype of a sample, we determined the total number of 20AARs retained across both alleles at tumour initiation. Truncating mutations in regions $R_1$, $R_2$ and $R_3$ are downstream of the last exon-exon junction of *APC* and therefore evade nonsense-mediated decay (NMD) [18]. Thus, we assume that frameshift and stop-gained mutations in regions $R_1$, $R_2$ and $R_3$ result in a translated protein with 1, 2 and 3 copies of 20AARs retained, respectively. Even though some stop-gained mutations of region $R_0$ are potentially targeted for NMD [18] (Figure 1b) these would result in 0 copies of 20AARs translated, hence we assume that truncating mutations in region $R_0$ result in 0 copies of 20AARs. Finally, under the assumption that whole-genome duplications occur after APC inactivation, the total number *X* of 20AARs expressed at initiation is independent of the WGD status. All together, we determined the total number *X* of 20AARs retained at the protein level using the mutation and copy-number status of the sample as follows:

- If a sample has a mutation in region $R_M$ in one allele and a mutation in region $R_N$ in the other allele, a total of *X*=*M*+*N* 20AARs are translated across both alleles.
- If upon a mutation in region $R_M$ the second hit is a copy-loss LOH, then the total number of 20AARs translated is *X*=*M*.
- If upon a mutation in region $R_M$ the second hit is a copy-neutral LOH, then the total number of 20AARs translated is *X*=2*M*.

**Classification by site.** The anatomical site of the tumour was defined corresponding to the position of the tumour at sampling. We classified samples occurring in the proximal colon (n=359) and those in the distal or rectum colon (n=620). The rest of samples were excluded from the site-specific analysis.

**Mutations in additional genes.** We considered additional the Wnt driver genes highlighted in van Ginkel *et al.* [55]. Presence of clonal driver mutations in other Wnt genes (AMER1, AXIN1, AXIN2, BCL9, BCL9L, CTNNB1, FBXW7, JUN, RNF43, SOX9, TCF7L2, ZNRF3, RSPO) as previously determined by Cornish *et al.* [3].

**MSI and *POLE* analysis.** For the hypermutant tumour analysis, we considered tumours with high microsatellite instability (MSI), as classified by [3]. The 100kGP CRC cohort includes 360 MSI tumours, out of which 110 had biallelic *APC* mutations (30.55%). Out of the 110, 15 samples were excluded because the mutations were not truncating mutations upstream SAMP. We also identified 18 samples with *POLE* mutations established as pathogenic [58]. These were further checked for hyper and ultra mutant (>100 mutations/megabase). All *POLE* samples had biallelic *APC* mutations (100%).

**Mutational signature of MSI samples.** The mean exposure vectors for single-base substitution and indel signatures in MSI samples was calculated using the signature analysis performed by Cornish *et al.* [3].

**Data accessibility.** Genomics England data is available for users of regecip. Summary tables including the number of samples with each *APC* genotype for different cohorts and the scripts required to perform the main analysis are available at https://github.com/xellbrunet/APC_Analysis_Public.

*1.2 cBioPortal and GENIE*

The same analysis outlined above was performed to classify *APC*-mutant tumours in an independent cohort combining public data accessed through cGenieBioPortal [4,5] (Supplementary Table 1). Primary tumours with two pathogenic mutations in *APC* and no copy-number alterations at *APC* locus other than WGD were considered, resulting in n=1,305 samples. The data and scripts used are available at https://github.com/xellbrunet/APC_Analysis_Public.

*1.3 FAP Patients*

To compare the distribution of the somatic hit on tumours in FAP patients, we used three published data-sets: [59] and [15,16] collected 93 adenomas, 55 adenomas and 86 cancer samples from 53, 18 and 23 FAP patients, respectively, and recorded the position and type of the germline mutation as well as any other mutations or loss of heterozygosity on *APC*. We further classified samples by *APC* genotype as outlined in section M1.1. As we could not distinguish between CN-LOH and CL-LOH, these were both classified as LOH. Characteristics of the cohorts can be found in the source publications. The data and scripts used are available at https://github.com/xellbrunet/APC_Analysis_Public.

## M2. Mathematical model of CRC initiation

We propose a mathematical model of *APC*-driven CRC initiation to estimate the relative probability of a given *APC* genotype resulting in tumour progression, where we define the *APC* genotype of a cell according to the type of the two inactivating alterations. We first outline the model, before detailing how it is parameterised and used for inference.

*2.1 Defining the CRC progression probability of APC genotypes*

We model the accumulation of *APC* mutations in colonic stem cells, with the following mutation types considered: SNV, small indels, copy-neutral LOH, and copy-loss LOH. All colonic stem cells acquire APC inactivating mutations at estimable, low mutation rates per year (estimates given below). Once a stem cell acquires a first *APC* mutation, we assume that the lineage of this cell fixes within the colon crypt by drift with a given probability which is constant over all mono-allelic *APC* mutant cells. Stem cells in single mutant crypts continue to accumulate *APC* mutations, leading to bi-allelic APC inactivation. We separate our explanation of the model into those pertaining to mutation, and selection.

**Mutation**. To model the mutational paths to bi-allelic APC inactivation, and to track the number of intact 20AAR domains, we introduce the following labelling, let: $[W, W]$ be stem cells without an *APC* mutation on either allele; types $\{0, 1, 2, 3\}$ denote truncating mutations in regions $R_0$, $R_1$, $R_2$ or $R_3$, respectively; type $-$, which denote copy loss of an allele (CL-LOH); and type $x2$, which denotes copy neutral loss of an allele (CN-LOH). Let $m_i$ be the probability that when a mutation occurs on an allele is of type $i$, where $i \in \{0, 1, 2, 3, -, x2\}$. Thus, given that a double wild type colonic cell gets a first *APC* mutation, it becomes of type $i$ with probability $m_i$. Our inference procedure to estimate the numerical values of $m_i$ is detailed below in 'Mutation and genotype probabilities'. We label the cell with a single mutation of type $i$ by $[i, W]$. Cells with a single *APC* mutation can acquire a mutation on the remaining wild type allele. If this mutation occurs, it is of type $j$ with probability $m_j$, where again $j \in \{0, 1, 2, 3, -, x2\}$. Hence, given that a cell has accumulated two *APC* mutations, these are of type $i$ and type $j$ respectively, with probability $m_i m_j$. We label the biallelic mutant cell by $[i, j]$. This two-step process can be illustrated by

$$[W,W] \xrightarrow{m_i} [i,W] \xrightarrow{m_j} [i,j]$$

where the probability of the path is $m_i m_j$. In principle, arbitrarily long mutation paths leading to APC inactivation exist. However, due to the mutation rates being small ($\mu_{APC} = 6.22 \cdot 10^{-6}$, Methods), long paths are unlikely, and so we consider only mutation paths of length 2. Moreover, we disregard: CN-LOH as the first mutation event; CN-LOH following CL-LOH; and, double CL-LOH, i.e. [-,-] as this genotype is unobserved in cancer data.

We were unable to distinguish the temporal order of mutations in sequence data, apart from genotypes resulting from CN-LOH. Thus, we introduce the further the genotype label (*M,N*) where $M \in \{0, 1, 2, 3\}$ is the region containing the furthest upstream truncating mutation, and $N \in \{0, 1, 2, 3, -, x2\}$ is either: the region containing the other truncating mutation; " $-$ " for CL-LOH; or "$x2$" for CN-LOH. Thus, e.g., both [1,2] and [2,1] map to (1,2). The probability of any (*M,N*) genotype can be written in terms of $m_i m_j$, and we normalise these probabilities such that they sum to one. Specifically, the probability that that a cell with bi-allelic inactivation of *APC* has genotype (*M,N*) is:

$$m_{(M,N)} \propto K\, m_M\, m_N, \text{ (Equation M1)}$$

where K=1 if $N \in \{M, x2\}$ and K=2 otherwise due to genotype labelling and the proportionality is due to normalising. Details concerning inferring the numerical values of $m_{(M,N)}$ using mutational signatures are given in methods 2.2.

**Selection.** We suppose that cells with biallelic APC inactivation can stochastically progress to CRC, with a cancer progression probability that depends on the *APC* genotype (*M,N*), which we denote $p_{(M,N)}$. Hence, combining the mutational probabilities of acquiring *APC* genotypes with the progression probabilities, each CRC has genotype (*M,N*) with probability

$$f_{(M,N)} = \frac{m_{(M,N)} p_{(M,N)}}{\sum_{(i,j)} m_{(i,j)} p_{(i,j)}}.$$

Therefore, the progression probability of genotype (*M,N*) is

$$p_{(M,N)} = C \frac{f_{(M,N)}}{m_{(M,N)}}, \text{ (Equation M2)}$$

where $C = \sum_{i,j} m_{(i,j)} p_{(i,j)}$ is a constant that does not depend on the type. As we aim to assess the relative oncogenic effect of different *APC* genotypes, we primarily focus on the relative progression probability of different *APC* genotypes

$$\widetilde{p}_{(M,N)} = \frac{p_{(M,N)}}{\sum_{i,j} p_{(i,j)}} = \frac{f_{(M,N)}/m_{(M,N)}}{\sum_{i,j} f_{(i,j)}/m_{(i,j)}}, \text{ (Equation M3)}$$

$f_{(M,N)}$ can be estimated from the frequency of CRCs with genotype ($M,N$) in cohorts of *APC*-mutant CRCs, whilst the probability of biallelic APC inactivation being of a given genotype $m_{(M,N)}$ is estimated by integrating mutational signature data with the genomic sequence of *APC* (see 'Mutation and genotype probabilities' section below).

In both 100kGP and cBioPortal cohorts, we found that the total number of 20AARs explains a considerable degree of variability in the relative progression probabilities between genotypes (Figure 3), supporting a model in which the total number of 20AARs across both alleles determines APC activity. This motivated classifying samples by the total number of 20AARs. The progression probabilities of cells with a given total number of 20AARs can be similarly calculated. Let $X$ denote the number of 20AARs retained in a cell with genotype ($M, N$) with $M \in \{0, 1, 2, 3\}$, where $X = M + N$ if $N \in \{0, 1, 2, 3\}$, $X = 2M$ if $N =$ "x2" and $X = M$ if $N = $ " $-$ ". To estimate the probability of a genotype occurring neutrally retaining $X$ copies of 20AARs, we sum of the probabilities of genotypes that result in $X$ retained 20AARs,

$$m_X = \sum_{(i,j):X} m_{(i,j)},$$

and similarly for the probability of a CRC with $X$ retained 20AARs,

$$f_X = \sum_{(i,j):X} f_{(i,j)}$$

which is estimated from the frequency of CRCs with $X$ 20AARs in the cohort of interest, as outlined in 1.1. The relative progression probability of $X$ retained 20AARs is given by

$$\widetilde{p}_X = \frac{p_X}{\sum_{y=0}^{6} p_y} = \frac{f_X/m_X}{\sum_{y=0}^{6} f_y/m_y} \text{ (Equation M4)}$$

**Comparing tumour subtypes.** We are interested in comparing different types of tumours, e.g, tumours in the distal or proximal colon or tumours with additional driver mutations. Let $\widetilde{p}_{x,A}$ denote the relative progression probability of $X$ retained 20AARs calculated using the mutational processes and genotype frequencies of a subset A of tumours. We define the `progression-weighted mean 20AARs number' of subset A as

$$\sum_{x=0}^{6} x \widetilde{p}_{x,A} \text{ (Equation M5)}$$

and compare disjoint subtypes of tumours $A$ and $B$ by computing the difference in the mean number of 20AAR,

$$\Delta_{A-B} := \sum_{x=0}^{6} x \widetilde{p}_{x,A} - \sum_{x=0}^{6} x \widetilde{p}_{x,B}. \text{ (Equation M6)}$$

**Progression probabilities in FAP patients.** FAP patients have germline mutations in APC, hence tumour progression only requires a somatic mutation on the non-mutated allele. Thus, under the model outlined above, for patients with germline mutation in region $R_M$, the probability that, given that they develop CRC, this has APC genotype (M,N) is

$$m_N p_{(M,L)} / (\sum_{i=0,1,2,3} m_i p_{(M,i)} + m_- p_{(M,-)} + m_{x2} p_{(M,x2)}). \quad \text{(Equation M7)}$$

## 2.2 Estimating mutation and genotype probabilities

In this section we estimate the probability that cells with bi-allelelic APC inactivation have genotype (M,N) under mutational processes alone, $m_{(M,N)}$. As the common factor of $m_0$ cancels in Eq. (!!! eqn for $m_{(M,N)}$) it is enough to estimate $m_i/m_0$ for $i \in \{0, 1, 2, 3, -, x2\}$. The data and scripts used are available at https://github.com/xellbrunet/APC_Analysis_Public.

**Truncating mutations.** We first estimate the probability a truncating mutation falls in a given region, that is $m_i/(\sum_j m_j)$. Let $p_{\text{stop-gain}}$ be the probability a new truncating mutation is a stop-gain, and $p_{\text{frameshift}}$ be defined analogously for frameshifts. To estimate these, we note the ratio of SBS to indels in healthy crypts was found to be 24:1 [30]. So by considering the number of single-base substitutions that can result in a stop codon upstream the SAMP repeat, we find that (737)/(3*4717) ~ 5.2% of SBS result in a stop-gained mutation whilst, by finding the total exposure of indel classes called in mutational signature analysis from healthy colon crypts [30] that disrupt the reading frame, we estimate that $\approx 88\%$ of indels result in a frameshift. Therefore, we estimate that the ratio of stop-gained to frameshifts in *APC* in healthy tissue is approximately 4:3, which is similar to the observed ratio in *APC* in the 100kGP MSS cohort, 1194/712 ~ 1.67. Thus we estimate $p_{\text{stop-gain}}$ = 4/7, with and $p_{\text{frameshift}}$ = 3/7.

We integrated SBS and indel mutational signature data with the genomic sequence of *APC* to estimate the probability that, when a stop-gain, or frameshift, mutation occurs, that it falls in region *i*, considering only mutational processes. We adopted the COSMIC V3 [60] mutation classes for SBS and indels but we omitted micro-homology ID classes, which account for <5% of IDs observed in healthy colonic crypts [30]. Separately for SBS and ID, we estimated P(new mutation is of class *x* and occurs at loci *i* in *APC*)=P(mutation occurs at loci *i* | mutation class *x*)*P(mutation class *x*). The term P(mutation occurs at loci *i* | mutation class *x*) is equal to 0 if loci *i* is not compliant with mutation class *x*; else it is equal to the number of *x*-compliant loci within *APC* (considering both complementary strands).

We estimate P(mutation class x) separately for both SBS and ID, but with a common procedure: We used signature data reported from sequencing healthy colonic crypts [30], and included only the ubiquitously observed SBS and ID signatures that were present in over 85% of crypts, which for SBS were SBS1, SBS5, SBS18, and for indels were ID1, ID2 and ID5. For SBS and ID separately: the relevant exposure vectors for each crypt were normalised such that the contributions from the ubiquitous signatures summed to 1; before averaging over crypt samples to create an average normalised exposure vector. The average normalised exposure vectors were used to weight the ubiquitously observed signatures, offering representative location-specific combined signatures for both SBS, and ID, which provided an estimate of P(mutation class x).

We summed P(new mutation is of class *x* and occurs at loci *i* in *APC*) for each stop-gained SBS in each region, and normalised to give the probability a new stop-gained occurs in a given region, and carried out the analogous procedure for frameshifts. Weighting these region probabilities by $p_{\text{stop-gain}}$ and $p_{\text{frameshift}}$ and summing ultimately provided the probability that, when a truncating mutation occurs within *APC*, it falls in each region. Taking the ratios of the *i*th region to region 0 we get $m_i/m_0$ for $i \in \{0, 1, 2, 3\}$.

**Copy number alterations.** To estimate the probability of copy number alterations relative to $m_0$ we assumed that genotypes (0,0), (0,-) and (0,x2) have the same progression probability, since they all result in complete loss of *APC* function. Then, using Equation M2, we have that

$$p_{(0,0)} = C \frac{f_{(0,0)}}{m_{(0,0)}} = C \frac{f_{(0,-)}}{m_{(0,-)}} = C \frac{f_{(0,x2)}}{m_{(0,x2)}}$$

Hence, the probabilities $m_{(0,-)}$, $m_{(0,x2)}$ can be obtained from the ratios of the probabilities of CRCs with the corresponding genotypes, which we estimate as the frequencies of cancers with the given genotype from cohort sequence data to obtain

$$\frac{m_{(0,-)}}{m_{(0,0)}} = \frac{2 m_-}{m_0} = \frac{f_{(0,-)}}{f_{(0,0)}} = 1.86 \quad \text{and} \quad \frac{m_{(0,x2)}}{m_{(0,0)}} = \frac{m_{x2}}{m_0} = \frac{f_{(0,x2)}}{f_{(0,0)}} = 1.43, \text{ (Equation M8-9)}$$

where we used Eq. M1 to relate the $m_{(M,N)}$ and $m_i$ terms. Thus, we again obtain $m_i/m_0$ for $i \in \{-, x2\}$. With cohort-specific estimates for $m_i/m_0$ for $i \in \{0, 1, 2, 3, -, x2\}$, we can calculate the mutation probabilities $m_{(M,N)}$ for all *APC* genotypes.

**Proximal-distal comparison.** To estimate the mutation probabilities of *APC* genotypes in proximal versus distal MSS cancers, we performed the analysis outlined above, parametrizing the model using site-specific signature exposure vectors of healthy crypts. Signature data were stratified by proximal colon (reported as 'right') and distal colon (reported as 'left') to calculate anatomical-site specific mutation probabilities $m_i$ for i=0,1,2,3. The rest of the parameters were kept the same.

**Hypermutant cancers.** The same analysis was performed to estimate the mutation probabilities of *APC* genotypes in POLE-deficient and MSI tumours. To obtain a combined signature exposure vector for POLE, we used signature data from individuals with germline POLE mutations [48]. For MSI, we used the signatures present in >85% of CRCs of the 100kGP (Supplementary Tables 3 and 4), determined by [3]. For *POLE*, since none of the *POLE*- deficient CRCs in our cohort had frameshift mutations in *APC*, we set $p_{\text{frameshift}}$ = 0. In the case of MSI tumours, the ratio of SBS to indels is 10:1 in the 100kGP cohort, hence we estimate the ratio of stop-gained to frameshifts in *APC* is 5:9, and so $p_{\text{stop-gain}}$ = 4/14, and $p_{\text{frameshift}}$ = 9/14. We also excluded samples with copy-number alterations, which are much less frequent in hypermutant tumours, thus in Equation M1, $m_{x2}=m_-=0$.

**Mutation rates per year.** We use the analysis above to estimate per year mutation rates, assuming that mutations occur at a constant rate in cells. Lee-Six et al (2019) estimate an average of 43.6 SBSs per year in healthy colonic stem cells, which corresponds to a rate of $\mu_{sbs} \approx 1.45 \cdot 10^{-8}$ substitutions per base-pair per year. The four considered *APC* regions comprise 4717 bases, and, as outlined above, only ~5.2% of possible SBSs within these regions are expected to be stop-gained mutations. Further, we expect a ratio of stop-gained to frameshifts of approximately 4:3. Hence, truncating mutations occur within the *APC* regions at a per-year rate of $\mu_{APC} = 4717 \cdot \mu_{sbs} \, 0.052 \cdot (1+\sfrac{3}{4}) = 6.22 \cdot 10^{-6}$. We can obtain the rate of truncating mutations at region $R_0$, $\mu_0 = \mu_{APC} m_0 = 5.28 \cdot 10^{-6}$. The frequencies of tumours with complete APC loss with genotypes (0,0), (0,-) and (0,x2) in the 100kGP cohort are $f_{(0,-)} = 0.44$, $f_{(0,-)} = 0.33$ Finally, from equations M7-8, the per year rates of copy-loss LOH and copy-neutral LOH in *APC* in healthy crypts are given by

$$\mu_- = \frac{f_{(0,-)}}{f_{(0,0)}} \cdot \frac{\mu_0}{2} = 4.72 \cdot 10^{-6} \text{ and } \mu_{x2} = \frac{f_{(0,x2)}}{f_{(0,0)}} \cdot \mu_0 = 7.18 \cdot 10^{-6}, \text{ respectively.}$$

*M3. Statistical analysis*

Statistical analysis was performed using Python. Standard statistical tests were performed and are described in the main text and figure legends, with confidence level 95% unless otherwise stated. Bootstrapping was performed using 1,000 iterations.

The following tests were designed to test the competing models for APC inactivation. For the 'Uniform risk' model, we reject the hypothesis that all genotypes have the same progression probability if the 95% confidence intervals of the progression probabilities are non-overlapping. For the 'Maximal APC loss implies maximal CRC risk model' and the 'Just-right model', we calculate the 95% confidence intervals of the mode of the distribution of relative progression probabilities using bootstrapping. We reject the hypothesis that maximal APC loss provides maximal CRC risk if '0' 20AARs is not in the confidence interval.

To determine if there were statistically significant differences in the progression probabilities between tumours with additional Wnt mutators correcting for the effect of anatomical site, we took the weighted average of the difference in progression-weighted mean 20AARs, $\Delta_{mutant-WT}$, conditioned on the anatomical site of the tumours. If the 95% confidence intervals of the estimator (obtained using bootstrapping) contains 0, we conclude that there was no statistically significant effect. We excluded AXIN1, AXIN2 and JUN as there were not enough samples to perform the site-correction (Supplementary Figure 2).

**Data availability.**

All data required to reproduce the mutation signature analysis, the main analysis using cBioPortal data and the FAP Analysis are available on GitHub https://github.com/xellbrunet/APC_Analysis_Public.
For 100kGP data, summary tables are provided on https://github.com/xellbrunet/APC_Analysis_Public. Full data is available to users of the 100kGP Genomics England portal and is found under re_gecip/cancer_colorectal/xellbrunet/APC_Analysis_Public. A csv file containing all data needed to reproduce the analysis of 100kGP data can be found in re_gecip/cancer_colorectal/xellbrunet/APC_Analysis_Public/all_data/APC_merged.csv. This combines summary information obtained by [3], as well as analysis performed for this study in particular.The following information can be found for each sample included:

- Origin files paths: the paths to the original tumour and germline bam files (filename_germline_bam, filename_snv_indel, filename_tumour_bam) and copy-number (cna) files (filename_can, filename_sv), which were generated by [3] from the bam files using the Battenberg Algorithm.
- APC information: the total number and position of stop-gained and frameshift mutations in APC (obtained from bam files), the copy number (obtained from cna files), and the APC genotype and expected total number of 20AARs, determined as outlined above. The scripts used to recover the extract information from the bam and cna files can be found in re_gecip/cancer_colorectal/xellbrunet/APC_paper/scripts/APC_information.
- Wnt drivers: mutation status of Wnt pathway-related genes (AMER1, AXIN1, AXIN2, BCL9, BCL9L, CTNNB1, FBXW7, JUN, RNF43, SOX9, TCF7L2, ZNRF3, RSPO (fusions)), determined by [3] from the bam and cna files using IntOGen.
- Tumour subtype: subtype (MSS/MSI/POL), whole-genome duplications, purity and ploidy, determined by [3] from the bam and cna files.
- Signature data information: number of variants attributed to single-base substitution signatures and indel signatures, determined by Cornish from the bam files.
- Clinical information: age at sampling, sex, tumour type, tumour anatomical site, any therapy prior sampling, ethnicity, death status, accessible via the LabKey participant information platform of the 100kGP Genomics England.

## Code availability

Code is available on GitHub https://github.com/xellbrunet/APC_Analysis_Publication and can be run to reproduce the analysis for which open access data is provided. The scripts used for analysis within the Genomics England environment are available on GitHub but can only be run within the environment.

## Acknowledgements

We thank Sjoerd V. Beentjes for helpful discussions. M.D.N. is a cross-disciplinary postdoctoral fellow supported by funding from CRUK Brain Tumour Centre of Excellence Award (C157/A27589).

## Author contributions

M.B.G., M.D.N., T.A., I.T. designed the research; M.B.G., M.D.N., T.A. developed the mathematical model; M.B.G. and M.D.N. performed the analysis; I.S., N.F. provided biological insight. S.T. provided bioinformatic support; M.D.N., T.A., I.T. supervised the research; M.B.G., M.D.N., T.A., I.T. wrote the manuscript. All authors approved the final manuscript.

## Competing Interests

The authors declare no competing interests.

# Supplementary information for *Quantifying 'just-right' APC inactivation for colorectal cancer initiation*

# Supplementary Figures

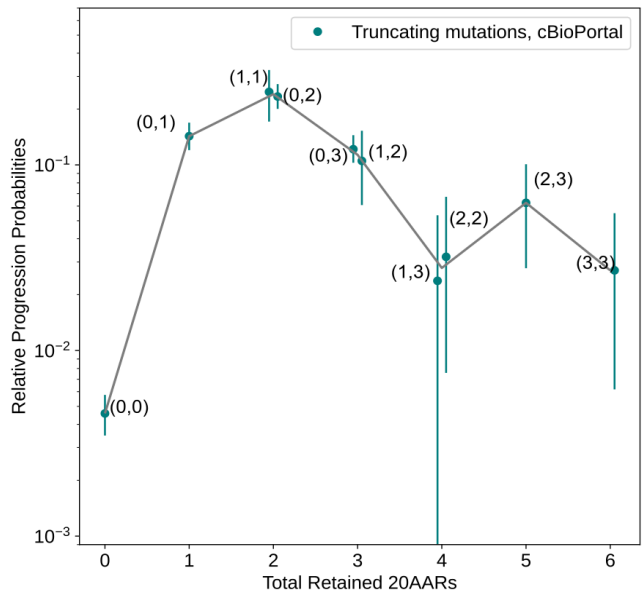

*Supplementary Figure 1. Progression probabilities of APC genotypes in cBioPortal Cohort.* The relative progression probability of different APC biallelic mutant genotypes, $\tilde{p}_{(M,N)}$, is plotted against the total number of 20AARs retained across both alleles. The frequencies of genotypes were calculated from sequence data of MSS primary CRCs in the cBioPortal cohort without copy-number alterations on APC (n=1,041, Methods). Whiskers for 95% confidence intervals (bootstrapping). The grey line is the weighted average of the progression probability over all genotypes which result in a given number of retained 20AARs.

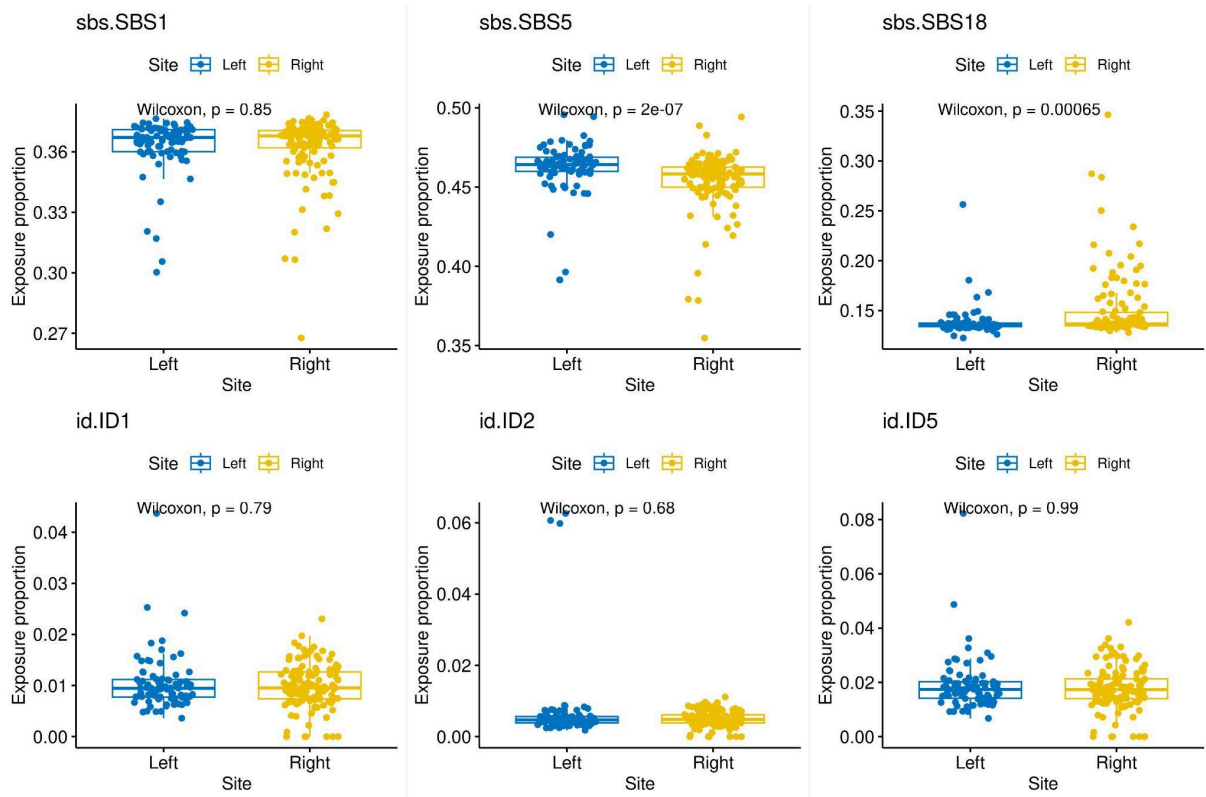

*Supplementary Figure 2. Site-specific differences in mutational signatures in the healthy colon.*
For mutational signatures observed ubiquitously in the healthy colon, we show the proportion of signature exposures, calculated by Lee-Six *et al* [2] for healthy colonic crypts labelled as Left colon (corresponding to distal) or Right colon (corresponding to proximal). Significant site-specific differences exist for SBS5 and SBS18, although the magnitude of the effect is relatively minor and so is unlikely to largely contribute to site-specificity of *APC* genotypes in CRCs.

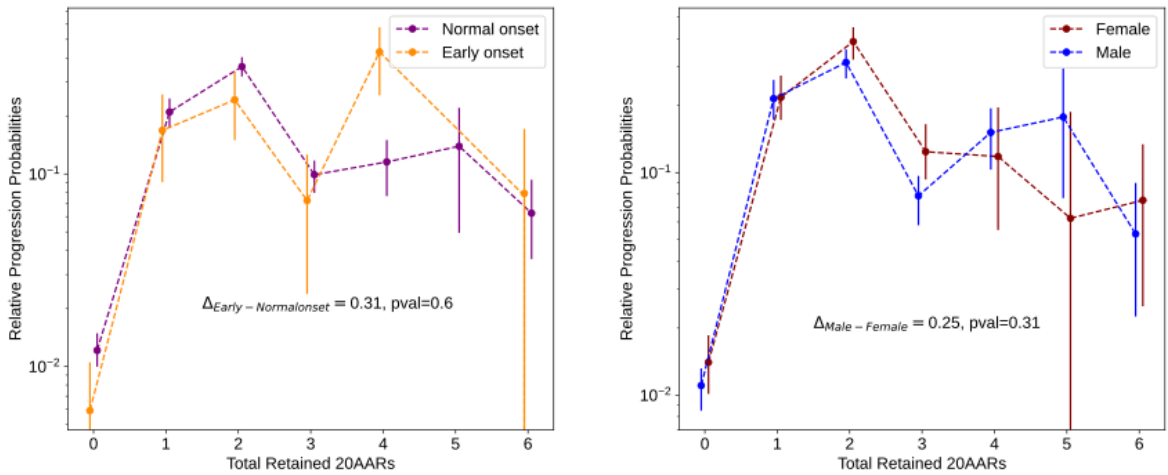

*Supplementary Figure 3. Differences in the progression probabilities by sex or age.*
The relative progression probability versus total number of 20AARs retained over both alleles, for (A) tumours in male (orange) versus female (purple) patients, and (B) in patients with early onset (<50 years old at resection, purple) versus normal onset (>50 years old at resection, yellow). Whiskers on points indicate 95% confidence intervals (bootstrapping). The difference in the progression-weighted mean 20AARs number is indicated by Δ (Methods). We find no statistically significant differences between tumours in male versus female patients (p=0.33, permutation test) nor in patients with early onset versus normal onset (p=0.73, permutation test).

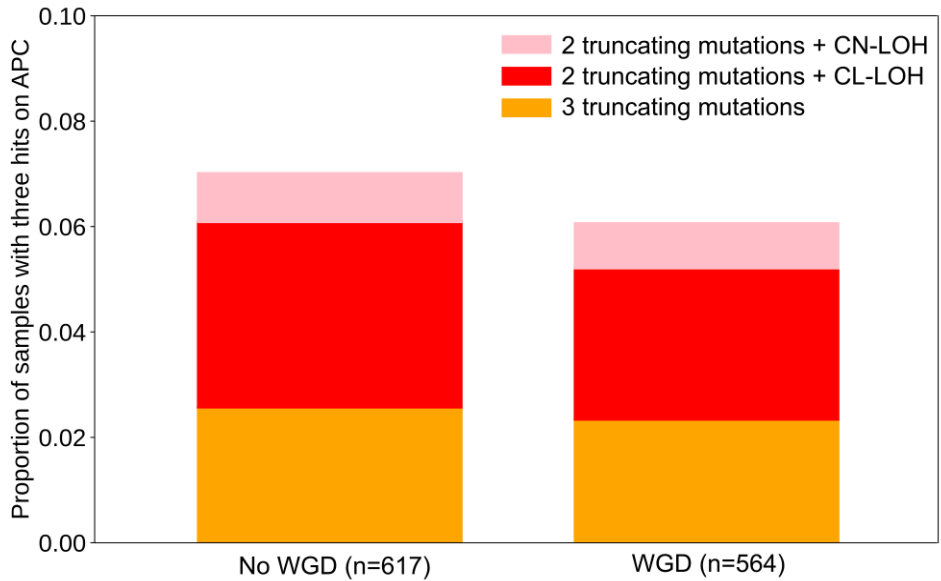

*Supplementary Figure 4. Third hits in APC in 100kGP* CRCs.
Proportion of APC-inactivated MSS primary CRCs in the 100kGP cohort (n=1,181) with evidence for third hits of different types, segregated by samples with no WGD (left) and with WGD (right), determined by Cornish et al (2022).The overall frequency of MSS CRCs with identifiable third hits in *APC* is 6.2%. Notably, in 100kGP the prevalence of third hits is not statistically different in samples with WGD p=0.58, chi2 statistic).

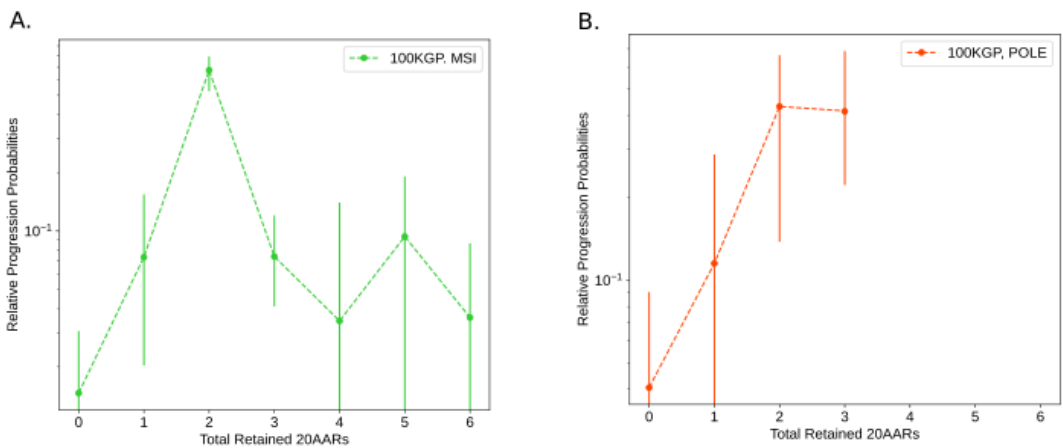

*Supplementary Figure 5. Progression probabilities in hypermutant CRCs.*
The relative progression probabilities by total number of 20AARs retained in MSI (A) and POLE-mutant CRCs (B) in the 100kGP cohort, as in main text Figure 6, but with bootstrapped confidence intervals.

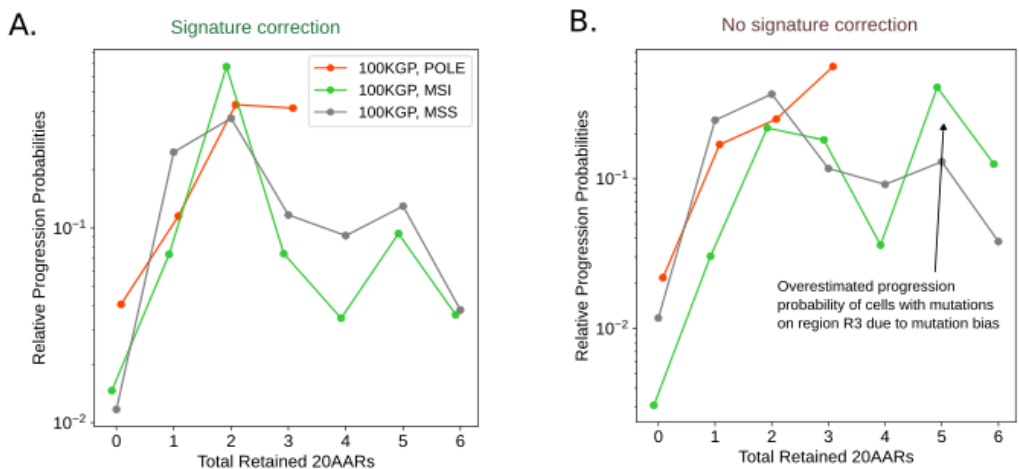

*Supplementary Figure 6. Signature correction in hypermutant CRCs.*
(A) The relative progression probabilities by total number of 20AARs retained in POLE-mutant, MSI and MSS CRCs 100kGP cohort, display overall agreement, which can be quantified by no significant differences in the progression-weighted mean number of 20AARs ($\Delta_{MSS-POLE}$=-0.22, 95% CI=[-0.62, 0.15], $\Delta_{MSS-MSI}$=-0.29, 95% CI=[-0.67, 0.08], bootstrapping). (B) As in (A) but without correcting for hypermutant mutational signatures. We see larger differences, which can be quantified by differences in the progression-weighted mean number of 20AARs which are statistically significant in the case of MSI tumours ($\Delta_{MSS-POLE,nc}$=-0.31, 95% CI=[-0.78, 0.2], $\Delta_{MSS-MSI,nc}$=-1.87, 95% CI=[-2.51, -0.81]). We highlight that the relative progression probability is larger for 5-6 total retained 20AARs for the MSI cohort without correction, due to an indel associated mutation bias to region 3.

## Supplementary Notes

**The rate of loss of heterozygosity (LOH)**

We estimated the rates of copy-loss and copy-neutral LOH at the *APC* locus of chromosome 5 (5q22.1–q22.3) in the healthy colon by assuming phenotypic equivalence between tumours with varied molecular causes of APC loss (Methods), resulting in estimates of $5.72*10^{-6}$ and $7.18\ 10^{-6}$/cell/year, respectively.

Paterson *et al*. [3] estimated the rate of LOH through a different strategy, obtaining a higher rate of $1.36*10^{-4}$/cell/year. Paterson *et al* used the ratio of MSI cancers with two inactivating mutations in *APC* to those with one inactivating mutation and one LOH event to be 1:7, as reported by Huang *et al*. [4] based on protein assays of *n*=55 CRCs. This ratio is in high discordance with the ratio in the 100kGP cohort of 10:1, based on a large number of whole-genome sequenced samples (n=2,023) [5]. Using the inference method of Paterson *et al.* with data from 100kGP we obtained a rate of LOH of $3.89*10^{-6}$/cell/year, which is comparable to our estimate.

## Supplementary Tables

**Supplementary Table 1. cBioPortal datasets**

| Study ID | Number of samples |
| --- | --- |
| coad_caseccc_2015 | 7 |
| coad_cptac_2019 | 42 |
| coad_silu_2022 | 123 |
| coadread_dfci_2016 | 105 |
| coadread_genentech | 8 |
| coadread_mskcc | 40 |
| coadread_mskresistance_2022 | 13 |
| coadread_tcga | 96 |
| crc_apc_impact_2020 | 149 |
| crc_dd_2022 | 24 |
| crc_nigerian_2020 | 11 |
| crc_public_genie_bpc | 593 |
| rectal_msk_2019 | 94 |
| Total primary CRC with two truncating mutations on APC | 1,305 |
| Total primary MSS CRC samples with two truncating mutations on APC and no copy number alterations at APC locus | 1,041 |

Supplementary Table 1. Public data accessed through cBioPortal [6,7] as of 1st of September 2023. Duplicated samples across different studies were removed. Primary tumours with two pathogenic mutations on APC and no copy-number alterations other than WGD were considered. The data and scripts used are available at [https://github.com/xellbrunet/APC_Public](https://github.com/xellbrunet/APC_Public).

## Supplementary Table 2. 100kGP APC CRC cohort

|  | Total | Primary MSS | Primary MSI | Primary POLE |
|---|---|---|---|---|
| All | 2023 | 1641 | 364 | 17 |
| APC mutant | 1499 | 1370 | 111 | 17 |
| APC biallelic inactivation | 1118 | 1037 | 64 | 17 |
| Supplementary Table 2. Number of samples considered from the 100kGP CRC cohort, corresponding to that analysed by (Cornish et al. 2022). | | | | |

## Supplementary Table 3. SBS signatures in MSI CRCs in 100kGP

| Signature | Proportion of samples | Mean exposure | Mean burden |
|---|---|---|---|
| SBS1 | 0.967 | 0.127 | 12140.95 |
| SBS5 | 0.981 | 0.33 | 32135.3 |
| SBS15 | 0.368 | 0.084 | 10280.01 |
| SBS26 | 0.296 | 0.073 | 9283.43 |
| SBS44 | 0.827 | 0.31 | 34619.09 |
| SBS57 | 0.329 | 0.07 | 8026.49 |

Supplementary Table 3. Single-base substitution mutational signatures present in >20% of MSI CRCs in the 100kGP cohort, determined by [5].

## Supplementary Table 4. ID signatures in MSI CRCs in 100kGP

| Signature | Proportion of samples | Mean exposure | Mean burden |
|---|---|---|---|
| ID1 | 0.989 | 0.134 | 18744.05 |
| ID2 | 1.000 | 0.866 | 123764.01 |

Supplementary Table 4. Insertion/deletion mutational signatures present in >20% of MSI CRCs in the 100kGP cohort, determined by [5].

# Supplementary Table 5. Signature analysis results

| APC Region | Stop-gained | | | | | Frameshifts | | | |
|---|---|---|---|---|---|---|---|---|---|
| | Healthy colon | Healthy right colon | Healthy left colon | POLE-deficient CRCs | MSI CRCs | Healthy colon | Healthy right colon | Healthy left colon | MSI CRCs |
| *R0* | 0.84876 | 0.84758 | 0.84904 | 0.78214 | 0.84827 | 0.73744 | 0.40667 | 0.50181 | 0.50368 |
| *R1* | 0.05476 | 0.05503 | 0.05438 | 0.08898 | 0.06335 | 0.06454 | 0.03586 | 0.04435 | 0.04452 |
| *R2* | 0.05457 | 0.05465 | 0.05444 | 0.04064 | 0.05034 | 0.08539 | 0.03714 | 0.04542 | 0.04558 |
| *R3* | 0.04250 | 0.04274 | 0.04215 | 0.08825 | 0.03804 | 0.11263 | 0.52032 | 0.40843 | 0.40622 |

Supplementary Table 5. The relative proportion of stop-gained and frameshifts expected to fall in different regions of *APC*, estimated by considering the ubiquitous mutational signatures found in healthy colon crypts [2], crypts with POLE mutations [8] and MSI CRCs in the 100kGP cohort (Methods). THese are used to estimate the mutation probabilities of different APC genotypes.

# Supplementary Table 6. APC genotype mapping

| Protein position of clonal frameshift or stop-gained mutations upstream codon 1569, with i<j<k | Copy number at APC locus | WGD | Inferred mutant type at initiation | Inferred APC genotype at initiation M: region of position i N: region of position j | n | Comments |
|---|---|---|---|---|---|---|
| i,j | | [1,1] | False | Bi-allelic mutant | (M,N) | 232 | |
| i,j | Variants need to be different | [a,b] , b>0 | True | Bi-allelic mutant | (M,N) | 187 | |
| i | | [2,0] | False | Copy-neutral LOH | (M, x2) | 104 | |
| i | | [1,0] | False | Copy-loss LOH | (M, 0) | 127 | |
| i | | [a,0], a>2 | True | Copy-neutral LOH | (M, x2) | 102 | |
| j | | [1,0], [2,0] | True | Copy-loss LOH | (M, -) | 150 | |
| i,j,k | Consider only the two most upstream mutations | [1,1] | False | Bi-allelic mutant | (M,N) | 16 | Three hits |
| i,j | Consider only the most upstream mutation | [2,0] | False | Copy-neutral LOH | (M, x2) | 1 | Three hits |
| i,j | Consider only the most upstream mutation | [1,0] | False | Copy-loss LOH | (M, -) | 5 | Three hits |
| i,j,k | Consider only the two most upstream mutations | [1,1] | True | Bi-allelic mutant | (M,N) | 13 | Three hits |
| i,j | Consider only the most | [2,0] | True | Copy-neutral LOH | (M, x2) | 5 | Three hits |
| | upstream mutation | | | | | | |
| i,j | Consider only the most upstream mutation | [1,0] | True | Copy-loss LOH | (M, -) | 16 | Three hits |
| i | | [a,b] with b>0 | Either | Single mutant | Excluded | 138 | |
| - | | Copy-number unknown or [0,0] | | | Excluded | 6 | |

Supplementary Table 6. Classification of *APC* genotypes at initiation. Mapping from *APC* sequence data acquired at tumour sample to *APC* genotype at initiation for all considered combinations of copy number, WGD and annotated variants.

Examples:

| Protein position of unique clonal frameshift or stop-gained mutations upstream codon 1569 | Copy number at APC locus | WGD | Inferred mutant type at initiation | Inferred APC genotype at initiation | Comments |
|---|---|---|---|---|---|
| [234 frameshift, 1311 stop-gain] | [1,1] | True | Bi-allelic mutant | (0,1) | |
| [1425 stop-gain, 1425 stop-gain] | [1,1] | True | Excluded | - | No evidence of biallelic loss |
| [1425 stop-gain] | [2,0] | Flase | Copy-neutral LOH | (2,x2) | |
| [1425 stop-gain] | [2,0] | True | Copy-loss LOH | (2,0) | |

## Supplementary Table 7. Model parameters

| Genotype $(M, N)$ | Total retained 20AARs | Mutation probability $m_{(M,N)}$ | Frequency in 100kGP $f_{(M,N)}$ | Relative CRC progression probability $\tilde{p}_{(M,N)}$ | 95 % CI for $\tilde{p}_{(M,N)}$ | |
|---|---|---|---|---|---|---|
| (0, 0) | 0 | 0.1733 | 0.0318 | 0.0036 | 0.0024 | 0.0050 |
| (1, 1) | 2 | 0.0009 | 0.0135 | 0.1899 | 0.0962 | 0.2868 |
| (2, 2) | 4 | 0.0012 | 0.0039 | 0.0425 | 0.0000 | 0.0941 |
| (3, 3) | 6 | 0.0014 | 0.0029 | 0.0373 | 0.0000 | 0.0841 |
| (0, 1) | 1 | 0.0255 | 0.1148 | 0.0870 | 0.0687 | 0.1070 |
| (1, 2) | 3 | 0.0022 | 0.0048 | 0.0573 | 0.0170 | 0.1011 |
| (1, 3) | 4 | 0.0023 | 0.0010 | 0.0158 | 0.0000 | 0.0406 |
| (0, 2) | 2 | 0.0293 | 0.1794 | 0.1164 | 0.0943 | 0.1386 |
| (2, 3) | 5 | 0.0027 | 0.0048 | 0.0411 | 0.0136 | 0.0755 |
| (0, 3) | 3 | 0.0314 | 0.0781 | 0.0531 | 0.0407 | 0.0660 |
| (0,-) | 0 | 0.3318 | 0.0723 | 0.0037 | 0.0027 | 0.0048 |
| (1,-) | 1 | 0.0244 | 0.0897 | 0.0654 | 0.0495 | 0.0833 |
| (2,-) | 2 | 0.0281 | 0.1389 | 0.1010 | 0.0817 | 0.1231 |
| (3,-) | 3 | 0.0301 | 0.0289 | 0.0168 | 0.0107 | 0.0236 |
| (0, x2) | 0 | 0.2525 | 0.0530 | 0.0037 | 0.0026 | 0.0048 |
| (1, x2) | 2 | 0.0186 | 0.1331 | 0.1263 | 0.1016 | 0.1548 |
| (2, x2) | 4 | 0.0214 | 0.0366 | 0.0289 | 0.0185 | 0.0399 |
| (3, x2) | 6 | 0.0229 | 0.0125 | 0.0104 | 0.0049 | 0.0163 |

Supplementary Table 7. Model parameters for APC-driven CRC initiation in the healthy colon. Mutation probabilities $m_{(M,N)}$ calculated using the mutational signatures ubiquitous to colonic crypts [2], Methods), frequencies of CRCs $f_{(M,N)}$ are calculated from the 100kGP cohort of MSS primary CRCs [5], relative progression probabilities $\tilde{p}_{(M,N)}$ are calculated using $m_{(M,N)}$ and $f_{(M,N)}$ as outlined in Methods. 95% CI obtained by bootstrapping (1,000 iterations).

## Supplementary Table 8. Odds ratio between APC and Wnt regulators

| Secondary Wnt | Odds ratio | p-value | Number of samples | | | |
|---|---|---|---|---|---|---|
| | | | APC only | Secondary Wnt only | Both | Neither |
| RNF43 | 0.019 | 3.938 E-24 | 1375 | 34 | 4 | 226 |
| ZNRF3 | 0.124 | 0.03 | 1377 | 3 | 2 | 257 |
| CTNNB1 | 0.152 | 1.464 E-05 | 1368 | 13 | 11 | 247 |
| AXIN1 | 0.281 | 0.180 | 1376 | 2 | 3 | 258 |
| AXIN2 | 0.788 | 0.590 | 1358 | 5 | 21 | 255 |
| BCL9L | 0.922 | 0.862 | 1325 | 11 | 54 | 249 |
| JUN | 1.037 | 1 | 1368 | 2 | 11 | 258 |
| FBXW7 | 1.441 | 0.155 | 1224 | 21 | 155 | 239 |
| BCL9 | 1.329 | 0.688 | 1337 | 6 | 42 | 254 |
| TCF7L2 | 2.355 | 0.001 | 1216 | 14 | 163 | 246 |
| SOX9 | 2.617 | 0.000720 | 1224 | 12 | 155 | 248 |
| AMER1 | 15.528 | 2.344 E-05 | 1301 | 1 | 78 | 259 |

Supplementary Table 8. Odds ratio and number of samples between the number of MSS primary CRCs with APC mutations and mutations in secondary Wnt drivers. P-values determined using Fisher's test. Presence of clonal driver mutations on other Wnt genes (AMER1, AXIN1, AXIN2, BCL9, BCL9L, CTNNB1, FBXW7, JUN, RNF43, SOX9, TCF7L2, ZNRF) as previously determined by Cornish *et al.* [5].

# Supplementary Table 9. Secondary Wnt variants

| Wnt regulator | Frameshift | Stopgain | Nonsynonymous SNV | Non frameshift insertion/deletion |
|---|---|---|---|---|
| AMER1 | 4 | 38 | 3 | |
| TCF7L2 | 44 | 13 | 35 | 2 |
| SOX9 | 50 | 21 | 10 | 3 |
| BCL9 | 13 | | 6 | 14 |
| FBXW7 | 5 | 16 | 72 | 1 |
| BCL9L | 10 | 9 | 11 | |
| JUN | 4 | 2 | 2 | |
| AXIN1 | 1 | | 7 | |
| AXIN2 | 9 | 3 | 10 | 1 |

Supplementary Table 9. Number and types of variants included in the analysis of secondary Wnt regulators as previously determined by Cornish *et al*. [5].

## Supplementary References